\documentclass[12pt,aps,amsmath,amssymb,superscriptaddress,pra]{revtex4-2}

\usepackage{amsmath,amssymb,amsthm}
\usepackage{graphicx}
\usepackage{dcolumn}
\usepackage{bbm}
\usepackage{color}
\usepackage{float}
\usepackage{caption}
\usepackage{subfig}
\usepackage{natbib}
\usepackage{indentfirst}
\usepackage{hyperref}

\newcommand{\safeincludegraphics}[2][]{%
  \IfFileExists{#2}{%
    \includegraphics[#1]{#2}%
  }{%
    \fbox{%
      \begin{minipage}[c][0.25\textheight][c]{0.75\textwidth}
      \centering
      Missing figure file: \texttt{#2}
      \end{minipage}
    }%
  }%
}

\newcommand{\bra}[1]{\langle #1|}
\newcommand{\ket}[1]{|#1\rangle}

\newcommand{\expect}[1]{\langle #1\rangle}

\begin{document}

\title{Correlation and Branching Mechanisms in the Quantum-to-Classical Transition of Interacting Systems}

\author{Bingyu Cui}
\email{bycui@cuhk.edu.cn}
\affiliation{
School of Science and Engineering,
The Chinese University of Hong Kong (Shenzhen),
Longgang, Shenzhen, Guangdong, 518172, P.R. China
}

\date{\today}

% ============================================================
% Abstract
% ============================================================
\begin{abstract}
\noindent We analyze the quantum-to-classical transition of interacting systems from the viewpoint of reduced dynamical closure. Bohmian trajectories are used as a diagnostic of the probability flow generated by the full wavefunction, rather than as an alternative set of quantum predictions. A closed classical or semiclassical description of a retained coordinate can emerge either because interaction-induced correlations are suppressed or because the observable of interest is insensitive to the correlations that remain. Three models are used to isolate distinct mechanisms. In a hard-core collision, the leading tail of a broad relative-coordinate wave packet reaches the boundary before its center and generates spatial nonseparability through reflection and interference. In the Rabi model, the mixed quantum-classical Ehrenfest approximation fails when spin-position covariances prevent the spin dynamics from being determined solely by the mean oscillator coordinate. In an internal-state-dependent force model, the total mean position obeys an exact Ehrenfest equation, while the wavefunction separates into branch-resolved pointer components; mean-level classicality therefore does not imply branch resolution. These examples show that the quantum-to-classical transition of interacting systems is controlled not only by wave-packet localization but also by whether and how interaction-induced correlations enter the reduced equations of motion.
\end{abstract}

\maketitle

% ============================================================
\section{Introduction}
\label{sec:introduction}
% ============================================================
The quantum-to-classical transition is a central problem in quantum foundations, measurement theory, decoherence, and semiclassical dynamics. In closed systems, classical behavior is commonly associated with wave-packet localization, action scales large compared with $\hbar$, and semiclassical propagation \cite{landau2013,dirac1981,Littlejohn1986}. In open systems, environmental decoherence provides an additional mechanism for suppressing interference between macroscopically distinct alternatives \cite{Zeh1970,Zurek2003,Schlosshauer2007}. For a single isolated degree of freedom, the emergence of classical mean motion can often be discussed through the Ehrenfest theorem \cite{Ehrenfest1927}, the mixed quantum-classical equations of motion \cite{Tully1990,Tully1998,Kapral1999,Kapral2006}, or through semiclassical wave-packet dynamics \cite{Littlejohn1986,Aguiar2005}.
%The quantum-to-classical transition is a central problem in quantum foundations, measurement theory, decoherence, and semiclassical dynamics. Classical behavior is often associated with narrow wave packets \cite{Demme2017,Kryukov2026}, large masses \cite{Sanz2008}, small effective de Broglie wavelengths \cite{DeBroglie1925}, environmental decoherence \cite{Zeh1970,Zurek2003,Schlosshauer2007}, or the formal limit $\hbar\rightarrow0$ \cite{landau2013,dirac1981}. For a single isolated degree of freedom, the emergence of classical motion can often be understood through the Ehrenfest theorem \cite{Ehrenfest1927,Cui2023b} or through semiclassical wave-packet dynamics \cite{Littlejohn1986,Lapa2019}.

For an interacting system, however, the classicality of a selected coordinate may not determined solely by the localization of its own wave packet. Once other degrees of freedom are integrated out, the selected coordinate is generally described by a reduced density matrix rather than by a pure one-particle wavefunction. Its dynamics need not be generated by a local effective potential. The key question is therefore whether the reduced dynamics closes: can the retained coordinate be described by a single classical or semiclassical trajectory, or do interaction-induced correlations remain dynamically relevant?

Mean-field Ehrenfest dynamics and its relation to surface-hopping and quantum-classical Liouville descriptions have been studied extensively \cite{Tully1990,Tully1998,Parandekar2006}. Its known limitations include incorrect equilibrium detailed balance, insufficient decoherence, and the inability of a single mean-field trajectory to represent wave-packet splitting associated with different quantum states. These shortcomings have motivated augmented Ehrenfest \cite{Subotnik2010,Subotnik2016}, multiconfigurational Ehrenfest \cite{Granucci2001,Shalashilin2009}, exact-factorization-based coupled-trajectory \cite{Min2015,Agostini2016,Gossel2018}, and phase-space mapping approaches \cite{Subotnik2013}. In the quantum Rabi model specifically, recent studies have formalized the Hamiltonian-level semiclassical limit \cite{TwyeffortIrish2022}, examined state-dependent convergence toward semiclassical dynamics \cite{Coleman2026}, and analyzed self-consistent mixed quantum-classical Rabi oscillations \cite{Hsieh2025}. The aim here is not to rediscover these generic shortcomings or to propose another corrected propagation algorithm. Instead, we ask whether the reduced equations for selected observables close, identify the specific correlations obstructing closure, and distinguish mean-level agreement from distributional concentration and spatial branch resolution.

In the following, we distinguish three related but inequivalent notions of classical behavior: (i) mean-level classicality (agreement between a quantum expectation value and a classical or semiclassical trajectory); (ii) distributional classicality (concentration of probability density around that trajectory); (iii) branch-level classicality (the formation of spatially distinguishable wave-packet components associated with different internal states \footnote{We use ``measurement-like branches'' rather than environmentally selected pointer states in the strict decoherence-theory sense.}). %spatially distinct components of the state to become sufficiently well resolved that they can be treated as separate alternatives. %The three models discussed in the article reveal that these notions need not emerge simultaneously.

Bohmian mechanics offers a particularly transparent framework for addressing this question \cite{Bohm1952,Bohm1952II,Holland1993,Durr1992}. In this formulation, particles possess definite positions guided by the many-body wavefunction. The wavefunction evolves according to the Schrödinger equation, with a matrix-valued Hamiltonian when internal (e.g. spin) degrees of freedom are present, while the particle trajectories are determined by the probability current. Individual Bohmian trajectories need not coincide with classical trajectories. Nevertheless, if the initial particle positions are sampled from the Born distribution, the trajectory ensemble reproduces the standard quantum probability distribution at all later times. In the infinite-sampling limit, its position averages are equal to the corresponding Born expectation values. Thus, Bohmian trajectories provide a useful diagnostic for comparing full quantum dynamics, ensemble-averaged motion, and classical or semiclassical predictions in a unified way.

The present work employs this trajectory perspective to analyze interacting systems during the quantum-to-classical transition. Bohmian trajectories are not used to propose a different set of quantum predictions. Instead, they are used as a diagnostic of the probability flow generated by the full wavefunction. The physical question is whether this flow can be reduced to closed classical or semiclassical dynamics for a selected coordinate. The answer depends on the correlations generated by the interaction. We consider initially separable states in which the spatial degrees of freedom are localized Gaussian wave packets, and vary the width while keeping a chosen mass-width scaling fixed, motivated by earlier analyses of center-of-mass states of large quantum systems \cite{Cui2023,Cui2026}. This also allows us to examine how the Bohmian ensemble becomes concentrated around classical or semiclassical paths, and how nonzero width effects generate departures from naive point-particle dynamics.

We study three examples. The first is a pair of equal-mass particles interacting through an impenetrable hard core \cite{Law2004,Harshman2008}. This model is simple enough to be solved by separating center-of-mass and relative coordinates, yet it already illustrates how an initially separable wavefunction becomes spatially correlated through scattering. The second example is the Rabi model \cite{Rabi1936} (see also Ref. \cite{,Xie2017,FornDaz2019} for modern analysis), in which a quantum harmonic oscillator is coupled to a two-level system. This model provides a natural setting for comparing full quantum dynamics with a mixed quantum-classical Ehrenfest approximation. The third example is an internal-state-dependent force (ISDF) model in which a spatial coordinate acts as a pointer subject to opposite forces depending on the internal state. Related internal-state-dependent models have appeared in discussions of quantum measurement \cite{vonNeumann1955,Maudlin1995} and in models of local vibrational dynamics in electronically excited states of molecules \cite{Cui2022,Cui2023b}. This model emphasizes the distinction between mean Ehrenfest motion and branch-resolved measurement-like dynamics.

The main message is that the agreement between Bohmian ensemble averages, Born expectation values, and classical or semiclassical dynamics is controlled not only by wave-packet localization but also by the correlations generated during the evolution. For sharply localized packets (with a fixed product of mass and initial spread square), the Bohmian ensemble becomes concentrated near the classical trajectory; in this limit, the trajectory ensemble becomes increasingly concentrated around the corresponding classical or semiclassical path. For broad (nonvanishing-width) packets, however, interactions can produce interference, entanglement, or branch-resolved structure. In such cases, an averaged trajectory may obey a classical-looking equation while still failing to characterize the underlying distribution or branch structure.

This paper is organized as follows. Section~\ref{sec:formalism} presents the formalism and diagnostics for interacting systems. Section~\ref{sec:results} scrutinizes the formalism in the hard-core model, the Rabi model, and the ISDF model. Section~\ref{sec:conclusion} summarizes the conclusions and discusses possible extensions.

% ============================================================
\section{Reduced Dynamical Closure and Bohmian Diagnostics}
\label{sec:formalism}
% ============================================================

\subsection{Bohmian probability flow and ensemble averages}

For simplicity, we first present the formalism for two nonrelativistic particles in one spatial dimension. The extension to more particles and higher dimensions is straightforward. The two-particle Schr\"{o}dinger equation is
\begin{equation}
    i\hbar\frac{\partial \Psi}{\partial t}
    =
    -\frac{\hbar^2}{2m_1}\frac{\partial^2\Psi}{\partial x_1^2}
    -\frac{\hbar^2}{2m_2}\frac{\partial^2\Psi}{\partial x_2^2}
    +V(x_1,x_2,t)\Psi,
    \label{eq:schrodinger_two_particle}
\end{equation}
where $m_1$ and $m_2$ are the particle masses, $V$ is the total potential, and $\Psi(x_1,x_2,t)$ is the two-body wavefunction.

%Writing the wavefunction in polar form,
%\begin{equation}
%    \Psi(\mathbf{x}_1,\mathbf{x}_2,t)
%    =
%    R(\mathbf{x}_1,\mathbf{x}_2,t)
%    \exp\left[
%    \frac{\ii}{\hbar}
%    S(\mathbf{x}_1,\mathbf{x}_2,t)
%    \right],
%    \label{eq:polar_form}
%\end{equation}
%and substituting Eq.~\eqref{eq:polar_form} into Eq.~\eqref{eq:schrodinger_two_particle}, one obtains the continuity equation
%\begin{equation}
%    \frac{\partial R^2}{\partial t}
%    +
%    \sum_{n=1}^2
%    \partial_n\cdot
%    \left(
%    \frac{R^2\partial_nS}{m_n}
%    \right)
%    =
%    0,
%    \label{eq:continuity}
%\end{equation}
%and the Hamilton-Jacobi-like equation
%\begin{equation}
%    \frac{\partial S}{\partial t}
%    +
%    \sum_{n=1}^2
%    \frac{(\partial_nS)^2}{2m_n}
%    +
%    V
%    +
%    Q
%    =
%    0.
%    \label{eq:quantum_HJ}
%\end{equation}
%The additional term
%\begin{equation}
%    Q(\mathbf{x}_1,\mathbf{x}_2,t)
%    =
%    -\frac{\hbar^2}{2R}
%    \sum_{n=1}^2
%    \frac{\partial_n^2R}{m_n}
%    \label{eq:quantum_potential}
%\end{equation}
%is the quantum potential. Since it depends on the amplitude of the full many-body wavefunction in configuration space, it can encode correlations between particles even when the classical potential is local or pairwise.

The Bohmian velocity of the $n$th ($n=1,2$) particle is defined by the guidance equation
%\begin{equation}
%    \mathbf{v}_n=\frac{\partial_nS}{m_n},
%    \label{eq:guidance_phase}
%\end{equation}
%or equivalently,
\begin{equation}
    v_n
    =
    \frac{j_n}{|\Psi|^2},
    \label{eq:guidance_current}
\end{equation}
where
\begin{equation}
    j_n
    =
    \frac{\hbar}{m_n}
    \operatorname{Im}
    \left(
    \Psi^*
    \frac{\partial \Psi}{\partial x_n}
    \right)
    \label{eq:single_current}
\end{equation}
is the probability current. %Taking a time derivative of Eq.~\eqref{eq:guidance_phase} and using Eq.~\eqref{eq:quantum_HJ}, one obtains the Newton-like form
%\begin{equation}
%    m_n
%    \frac{d ^2\mathbf{x}_n}{d  t^2}
%    =
%    -\partial_n
%    \left(
%    V+Q
%    \right).
%    \label{eq:newton_like}
%\end{equation}
Bohmian trajectories are then obtained by integrating the first-order guidance equation~\eqref{eq:guidance_current}.

Bohmian mechanics assigns a definite trajectory to each realization. The observable predictions, however, are obtained from an ensemble of initial configurations distributed according to the Born rule,
\begin{equation}
    \rho(x_1,x_2,0)
    =
    |\Psi(x_1,x_2,0)|^2.
    \label{eq:initial_born}
\end{equation}
The equivariance property guarantees that this distribution is preserved by the dynamics \cite{Durr1992,Oriols2019}:
\begin{equation}
    \rho(x_1,x_2,t)
    =
    |\Psi(x_1,x_2,t)|^2.
    \label{eq:equivariance}
\end{equation}
Therefore, in the infinite-ensemble limit, the Bohmian ensemble average of a position observable equals the corresponding Born expectation value. In practical Bohmian simulations, however, one instead compares the finite sample average
\begin{equation}
    \overline{x}^{(N)}_n(t)
    =
    \frac{1}{N}
    \sum_{k=1}^{N}
    x_{n,k}(t),
    \label{eq:sampleaverage}
\end{equation}
where $k$ labels the trajectory index and $N$ is the total number of trajectories,
with the Born expectation value
\begin{equation}
    \expect{x_n(t)}
    =
    \int
    x_n
    \rho(x_1,x_2,t)
    d x_1 d x_2.
    \label{eq:Bornaverage}
\end{equation}
 In general, the difference between Eqs. \eqref{eq:sampleaverage} and \eqref{eq:Bornaverage} decreases with increasing number of trajectories, provided the velocity field and wavefunction propagation are accurate. Any discrepancy between a numerically sampled Bohmian average and the Born expectation value is therefore a finite-sampling effect \cite{Cui2026b} or numerical error, rather than a physical difference between Bohmian mechanics and standard quantum mechanics.

\subsection{Spinor probability current}
For systems with spin or other internal degrees of freedom, the wavefunction is spinor-valued. For a two-component spinor with a single coordinate $x$,
\begin{equation}
    \boldsymbol{\Psi}(x,t)
    =
    \begin{pmatrix}
    \Psi_1(x,t)\\
    \Psi_2(x,t)
    \end{pmatrix},
    \label{eq:spinor_wavefunction}
\end{equation}
the probability density is
\begin{equation}
    \rho
    =
    \boldsymbol{\Psi}^\dagger
    \boldsymbol{\Psi}
    =
    \sum_{\alpha}
    |\Psi_\alpha|^2.
    \label{eq:spinor_density}
\end{equation}
For orthogonal internal states, the density is the incoherent sum over spinor components, as in Eq.~\eqref{eq:spinor_density}. No cross term appears in the spin-unresolved position density. Interference terms may appear only in a conditional spatial density obtained after postselection onto a coherent superposition of the internal states. The probability current is
\begin{equation}
    j
    =
    \frac{\hbar}{m}
    \operatorname{Im}
    \left[
    \Psi_1^*\frac{\partial\Psi_1}{\partial x}+\Psi_2^*\frac{\partial\Psi_2}{\partial x}
    \right],
    \label{eq:spinor_current}
\end{equation}
where $m$ is the mass and the Bohmian velocity field is again $v=j/\rho$.
%\begin{equation}
%    v
%    =
%    \frac{j}{\rho}.
%    \label{eq:spinor_guidance}
%\end{equation}

\subsection{Localized wave-packet limit}

The quantum-to-classical transition considered in this work is implemented through initially localized Gaussian packets. Such localized packets may arise, for example, in the center-of-mass description of a large quantum system \cite{Demme2017,Cui2023}. A useful one-dimensional Gaussian packet is
\begin{equation}
    \psi(x,0)
    =
    \frac{1}{(2\pi\sigma^2)^{1/4}}
    \exp\left[
    -\frac{(x-x_0)^2}{4\sigma^2}
    +
    \frac{i}{\hbar}p_0x
    \right],
    \label{eq:gaussian_packet_general}
\end{equation}
where $x_0$ is the packet center and $p_0$ is the (mean) momentum.
%Taking $\sigma\rightarrow0$, together with an appropriate mass-width scaling, produces a sharply localized distribution. %The quantum potential associated with the real amplitude of Eq.~\eqref{eq:gaussian_packet_general} at $t=0$ is
%\begin{equation}
%    Q(x,0)
%    =
%    -\frac{\hbar^2}{2m\sigma^2}
%    \left[
%    \frac{(x-x_0)^2}{4\sigma^2}
%    -
%    \frac{1}{2}
%    \right].
%    \label{eq:packetQ}
%\end{equation}
%For fixed $m\sigma^2$, the value of $Q$ at the packet center remains finite, and the quantum force $-\partial_x Q$ vanishes at $x=x_0$. 
%In the limit $\sigma\rightarrow0$, the Born distribution becomes concentrated near the packet center, so the ensemble becomes localized around the classical initial condition \cite{Cui2023}. 
For a free or constant-force Gaussian propagation with $\kappa\equiv m\sigma^2$ fixed, the propagated width is 
\begin{equation}
    \sigma_t
    =
    \sigma
    \sqrt{
    1+
    \left(
    \frac{\hbar t}{2\kappa}
    \right)^2
    },
    \label{eq:sigma_t}
\end{equation}
and therefore $\sigma_t\rightarrow0$ for every fixed time as $\sigma\rightarrow0$. For nonlinear or internally coupled systems, Eq. \eqref{eq:sigma_t} characterizes the initial scaling but does not determine the subsequent width; persistence of localization may then be examined dynamically. The corresponding classical limit can be understood as an asymptotic statement for a family of increasingly localized wave packets. In the model sequences considered below, the intended distributional classical limit is characterized by the weak convergence \footnote{In Bohmian language, this classical limit might be more easily interpreted: The Gaussian distribution with a vanishing width (i.e. delta-like distribution) assigns a unique spatial position (and momentum), at which the quantum force vanishes, leading to the deterministic Newtonian dynamics.},
\begin{equation}
    |\Psi(x_1,x_2,t)|^2
    \rightarrow
    \delta(x_1-x_1^{\rm cl}(t))
    \delta(x_2-x_2^{\rm cl}(t)).
    \label{eq:weak_classical_limit}
\end{equation}
The limiting classical coordinates obey
\begin{equation}
    \frac{d  x_n^{\rm cl}}{d  t}
    =
    \frac{p_n^{\rm cl}}{m_n},
    \qquad
    \frac{d  p_n^{\rm cl}}{d  t}
    =
    -\left.\frac{\partial V(x_1,x_2,t)}{\partial x_n}\right|_{x_1=x_1^{\rm cl}(t),x_2=x_2^{\rm cl}(t)},\quad n=1,2.
    \label{eq:classical_equations}
\end{equation}

\subsection{Classical closure and correlation corrections}

For a general interacting system, the exact Ehrenfest equation for the mean coordinate is
\begin{equation}
    m_n
    \frac{d ^2\expect{x_n}}{d t^2}
    =
    \expect{F_n(x_1,x_2,t)},
    \qquad
    F_n
    =
    -\frac{\partial V}{\partial x_n}.
    \label{eq:exact_ehrenfest_force}
\end{equation}
A closed classical description replaces the expectation value of the force by the force evaluated at the mean configuration,
\begin{equation}
    m_n
    \frac{d ^2\bar{x}_n}{d  t^2}
    \approx
    F_n(\bar{x}_1,\bar{x}_2,t),
    \qquad
    \bar{x}_n\equiv\expect{x_n}.
    \label{eq:classical_force_closure}
\end{equation}
The difference between Eqs.~\eqref{eq:exact_ehrenfest_force} and \eqref{eq:classical_force_closure} is controlled by fluctuations and correlations. To see this explicitly, %let $q_i$ denote all coordinate components collectively, $\bar q_i=\expect{q_i}$, and $\delta q_i=q_i-\bar q_i$. Here $q$ labels a Cartesian component of the force acting on one of the retained coordinates. 
expand the force component $F_n(x_1,x_2)$ from the mean coordinates $\mathbf{\bar x}=(\bar x_1,\bar x_2)$,
\begin{align}
    \expect{F_n(x_1,x_2)}
    &=F_n(\bar x_1, \bar x_2)
    +\left.\frac{1}{2}\frac{\partial^2 F_n}{\partial x_1^2}\right|_{\mathbf{\bar x}}\text{Var}(x_1)+\left.\frac{1}{2}\frac{\partial^2 F_n}{\partial x_2^2}\right|_{\mathbf{\bar x}}\text{Var}(x_2)\notag\\
    &+\left.\frac{\partial^2 F_n}{\partial x_1\partial x_2}\right|_{\mathbf{\bar x}}\text{Cov}(x_1,x_2)+\mathcal{O}(\expect{||\delta \mathbf{x}||^3}).
    \label{eq:force_cumulant_expansion}
\end{align}
Here $\delta x_n=x_n-\bar x_n$, $\text{Var}(x_n)=\langle \delta x_n^2\rangle$ are the variances and $\text{Cov}(x_1,x_2)=\expect{\delta x_1\delta x_2}$ is the intercoordinate covariance. The remainder is governed by third- and higher-order central moments. Thus, classical closure requires not only narrow marginal distributions but also negligible correlations and higher moments. The examples below illustrate three ways in which this closure may fail or become misleading. It is important to note that \eqref{eq:force_cumulant_expansion} requires the force to be sufficiently smooth over the region occupied by the wavefunction. It does not apply directly, for example, to the singular hard-core force, whose effect is instead handled through the boundary condition at  $x_1=x_2$.

\subsection{Reduced one-coordinate probability flow}

After eliminating the second coordinate, particle 1 is generally described by the reduced density matrix
\begin{equation}
    \rho_1(x,x';t)
    =
    \int
    \Psi(x,x_2,t)
    \Psi^*(x',x_2,t)
    d x_2 .
\end{equation}
The corresponding marginal density and current are
\begin{equation}
    \rho_1(x,t)=\rho_1(x,x;t),
\end{equation}
and
\begin{equation}
    J_1(x,t)
    =
    \int
    j_1(x,x_2,t)
    d x_2 .
\end{equation}
which obey the continuity equation,
\begin{equation}
    \frac{\partial \rho_1}{\partial t}
    +
    \frac{\partial J_1}{\partial x}
    =
    0.
\end{equation}
The reduced probability-flow velocity is therefore
\begin{equation}
    u_1(x,t)=\frac{J_1(x,t)}{\rho_1(x,t)}.
\end{equation}
In general, this reduced flow is not necessarily equivalent to the velocity defined through two-body density and current, cf. Eq. \eqref{eq:guidance_current}. In addition, it is not generated by a closed one-particle Schr\"{o}dinger equation with a local scalar potential. However, the reduced-flow ensemble mean also satisfies
\begin{equation}
    \expect{x_1(t)}=\int x\rho_1(x,t)dx,
\end{equation}
in agreement with the average position of the full two-body Bohmian ensemble.

\subsection{Diagnostics of nonclosure}
We will use three diagnostics to identify the mechanism by which classical closure succeeds or fails. First, to quantify the spatial nonseparability generated by the interaction for a pair of particles, we compute linear entropy in terms of $\rho_1$ \cite{Nielsen_Chuang_2010}, %the reduced one-coordinate density matrix by tracing over the second coordinate,
%\begin{equation}
 %   \rho_1(x_1,x_1';t)
%    =
%    \int
%    \Psi(x_1,x_2,t)
%    \Psi^*(x_1',x_2,t)
%    \,d  x_2 .
%    \label{eq:rho1_general_app}
%\end{equation}
%Since the hard-core wavefunction is zero in the forbidden region $x_1\leq x_2$, the reduced density matrix can equivalently be written as
%\begin{equation}
%    \rho_1(x,x';t)
%    =
%    \int_{-\infty}^{\min(x,x')}
%    \Psi(x,x_2,t)
%    \Psi^*(x',x_2,t)
%    \,d  x_2 .
%    \label{eq:rho1_hardcore_app}
%\end{equation}
\begin{equation}
    S_L(t)
    =
    1-\operatorname{Tr}\rho_1^2(t).
    \label{eq:linear_entropy_app}
\end{equation}
A separable pure state has $S_L=0$, while $S_L>0$ indicates that the reduced one-coordinate state is mixed due to spatial nonseparability. In the present treatment, particles of equal mass are taken to be distinguishable labeled subsystems. Accordingly, $S_L$ quantifies entanglement with respect to the $x_1|x_2$ particle partition.

If the particle carries a spin (say $1/2$), we monitor the failure of semiclassical closure through (raw) spin-position covariances, e.g.,
\begin{equation}
    C_{x\sigma_\alpha}(t)
    =
    \expect{x\sigma_\alpha}
    -
    \expect{x}\expect{\sigma_\alpha},
    \qquad
    \alpha=y,z.
\end{equation}
These covariances measure the extent to which the spin dynamics cannot be determined solely by the mean spatial coordinate.

Last, if the statistics of the particle position exhibits branching, we characterize the branch resolution by the separation, $\Delta x(t)$, between the (two) branch centers relative to their respective position uncertainties $\sigma_{1}(t),\sigma_{2}(t)$ at time $t$,
\begin{equation}
    R(t)
    =
    \frac{\Delta x(t)}{\sigma_{1}(t)+\sigma_{2}(t)}.
    \label{eq:branch_resolution}
\end{equation}
If the two branches have equal position spread $\sigma(t)$, then $R(t)=\Delta x(t)/2\sigma(t)$. When $R(t)\ll1$, the branches distributions strongly overlap, whereas $R(t)\gg1$ indicates well-resolved pointer branches.

Before proceeding to the detailed model analysis in the next section, we provide a brief summary of the correlation mechanisms and diagnostics in Table~\ref{tab:mechanisms}.

\begin{table*}[t]
\caption{Mechanisms controlling the quantum-to-classical transition.}
\label{tab:mechanisms}
\centering
\small
\setlength{\tabcolsep}{8pt}
\renewcommand{\arraystretch}{1.15}
\begin{tabular}{@{}llll@{}}
\hline
Model
&
\begin{tabular}[t]{@{}l@{}}
Naive classical\\
picture
\end{tabular}
&
Correlation mechanism
&
Diagnostic
\\
\hline
\begin{tabular}[t]{@{}l@{}}
Hard-core\\
collision
\end{tabular}
&
\begin{tabular}[t]{@{}l@{}}
Elastic exchange\\
of momenta.
\end{tabular}
&
\begin{tabular}[t]{@{}l@{}}
Reflection of the relative-coordinate wave packet\\
generates spatial nonseparability.
\end{tabular}
&
\begin{tabular}[t]{@{}l@{}}
$S_L(t)$, $\rho_1(t)$,\\ $J_1(x,t)$, $u_1(x,t)$.
\end{tabular}
\\[0.8em]
Rabi model
&
\begin{tabular}[t]{@{}l@{}}
Oscillator follows\\
an Ehrenfest trajectory.
\end{tabular}
&
\begin{tabular}[t]{@{}l@{}}
Spin-position covariances break\\
semiclassical closure.
\end{tabular}
&
\begin{tabular}[t]{@{}l@{}}
$\langle x(t)\rangle -x_{semi}(t)$,\\ $\xi_{\text{cov}}(t)$, $C_{x\sigma_{\alpha}}(t)$.
\end{tabular}
\\[0.8em]
\begin{tabular}[t]{@{}l@{}}
ISDF model
\end{tabular}
&
\begin{tabular}[t]{@{}l@{}}
Pointer follows the\\
mean force.
\end{tabular}
&
\begin{tabular}[t]{@{}l@{}}
Branches separate while the total\\
mean remains Ehrenfest.
\end{tabular}
&
\begin{tabular}[t]{@{}l@{}}
$n_g(x,t)$, $n_e(x,t),$\\ $R(t)$, $\mathcal{B}_\rho(t)$.
\end{tabular}
\\
\hline
\end{tabular}
\end{table*}

% ============================================================
\section{Three mechanisms in solvable models}
\label{sec:results}
% ============================================================

We now apply the above formalism to three interacting models. Unless otherwise stated, all quantities are reported in arbitrary units. Detailed numerical schemes are sketched in Appendix \ref{app:numerical}.

% ============================================================
\subsection{Hard-core collision: boundary-induced spatial nonseparability}
\label{subsec:hard_core}
% ============================================================
The first model consists of two labeled, distinguishable particles of equal mass interacting through an impenetrable hard core. They may be viewed, for example, as different species or as particles carrying distinct spectator labels. The allowed region is chosen as the ordering $x_1>x_2$, or
\begin{equation}
    V(x_1,x_2)
    =
    \begin{cases}
    0, & x_1>x_2,\\
    \infty, & x_1\leq x_2.
    \end{cases}
    \label{eq:hard_core_potential}
\end{equation}
The hard-core potential is equivalently imposed by the boundary condition
\begin{equation}
    \Psi(x,x,t)=0.
    \label{eq:hard_core_boundary}
\end{equation}

In classical dynamics, two equal-mass particles undergoing elastic collision exchange momenta. Quantum mechanically, the same process is elucidated by the reflection of the relative coordinate wavefunction from the boundary $x_1=x_2$. 

The initial state is chosen as a product of two Gaussian packets,
\begin{equation}
    \Psi(x_1,x_2,0)
    =
    \psi_1(x_1)\psi_2(x_2),
    \label{eq:hard_initial_product}
\end{equation}
where
\begin{equation}
    \psi_i(x_i)
    =
    \frac{1}{(2\pi\sigma_i^2)^{1/4}}
    \exp\left[
    -\frac{(x_i-x_{i,0})^2}{4\sigma_i^2}
    +
    \frac{i}{\hbar}p_ix_i
    \right],
    \qquad i=1,2.
    \label{eq:hard_initial_gaussians}
\end{equation}
The packet centers are $x_{i,0}$ and the (mean) momenta are $p_{i}$. The exact hard-core state is obtained by taking the odd image extension of the relative-coordinate wavefunction (the method of images) \cite{Schulman1981}. For initially well-separated packets, however, the image contribution is exponentially small, so the hard-core initial state is approximately equal to the product state in the allowed region.

For equal initial widths, $\sigma_1=\sigma_2=\sigma$, introduce the center of mass and relative coordinates, respectively,
\begin{equation}
    X=\frac{x_1+x_2}{2},
    \qquad
    r=x_1-x_2.
    \label{eq:CM_relative}
\end{equation}
The center-of-mass coordinate remains free, while the relative coordinate behaves as a particle on the half-line. By the method of images, the wavefunction can be written as
\begin{equation}
    \Psi(x_1,x_2,t)
    =
    \mathcal{N}
    \psi_{\rm CM}(X,t)
    \left[
    \phi_{\rm free}(r,t)
    -
    \phi_{\rm free}(-r,t)
    \right],
    \qquad r>0,
    \label{eq:hard_core_solution}
\end{equation}
where $\mathcal{N}$ is a normalization constant. The center-of-mass wave packet is
\begin{align}
    \psi_{\rm CM}(X,t)
    &=
    \frac{1}{(\pi\sigma^2)^{1/4}\sqrt{D(t)}}
    \exp\left[
    -\frac{(X-X_c(t))^2}{2\sigma^2D(t)}
    +
    \frac{i}{\hbar}(p_1+p_2)
    \left(
    X-\frac{(p_1+p_2)t}{4m}
    \right)
    \right],
    \label{eq:CM_packet}
\end{align}
and the free relative-coordinate packet is
\begin{align}
    \phi_{\rm free}(r,t)
    &=
    \frac{1}{(4\pi\sigma^2)^{1/4}\sqrt{D(t)}}
    \exp\left[
    -\frac{(r-r_c(t))^2}{8\sigma^2D(t)}
    +
    \frac{i}{\hbar}\frac{p_1-p_2}{2}
    \left(
    r-\frac{(p_1-p_2)t}{2m}
    \right)
    \right].
    \label{eq:relative_packet}
\end{align}
Here
\begin{align}
    D(t)
    &=
    1+\frac{i\hbar t}{2\kappa},
    \label{eq:D_factors}
\end{align}
is the complex spreading factor and
\begin{align}
\label{eq:oldcenters1}
    X_c(t)&=\frac{x_{1,0}+x_{2,0}}{2}+\frac{(p_1+p_2)t}{2m},
    \\
    r_c(t)&=x_{1,0}-x_{2,0}+\frac{(p_1-p_2)t}{m}.
    \label{eq:oldcenters2}
\end{align}
are packet centers at later times; $m$ is the mass.
Derivation details are provided in Appendix~\ref{app:hard_core_derivation}. In the main scaling sequence we fix both the initial velocities $v_n$ and $\kappa\equiv m\sigma^2$. Since $p_n=mv_n$, Eqs. \eqref{eq:oldcenters1} and \eqref{eq:oldcenters2} become
\begin{align}
\label{eq:centers1}
    X_c(t)&=\frac{x_{1,0}+x_{2,0}}{2}+\frac{(v_1+v_2)t}{2},
    \\
    r_c(t)&=x_{1,0}-x_{2,0}+(v_1-v_2)t.
    \label{eq:centers2}
\end{align}
Hence the packet centers and the classical collision time are identical for all values of $\sigma$. Moreover, $D(t)$ is common to the entire sequence. The time-dependent center-of-mass and relative-coordinate widths are therefore
\begin{align}
    \sigma_X(t)=\frac{\sigma}{\sqrt{2}}|D(t)|;\quad \sigma_r(t)=\sqrt{2}\sigma|D(t)|.
\end{align}
Thus, along this scaling sequence, decreasing $\sigma$ reduces the absolute spatial extent of both wave packets at every fixed time while leaving the classical centers, the collision time, and the dimensionless spreading factor unchanged. The scan isolates localization under the fixed-$\kappa$, fixed-velocity protocol, although the mass, momentum, and kinetic energy necessarily vary with $\sigma$. The exact mean of each particle can be calculated by integrating out the relative-coordinate, which is
\begin{equation}
    \expect{x_n(t)}=X_c(t)+(-1)^{n-1}\frac{1}{2}\expect{r(t)},
    \label{eq:meanx}
\end{equation}
where
\begin{equation}
    \expect{r(t)}=\frac{\int_0^\infty r|\phi_{free}(r,t)-\phi_{free}(-r,t)|^2dr}{\int_0^\infty |\phi_{free}(r,t)-\phi_{free}(-r,t)|^2dr}.
    \label{eq:meanr}
\end{equation}

\begin{figure}[!tp]
\centering
\safeincludegraphics[width=0.8\textwidth]{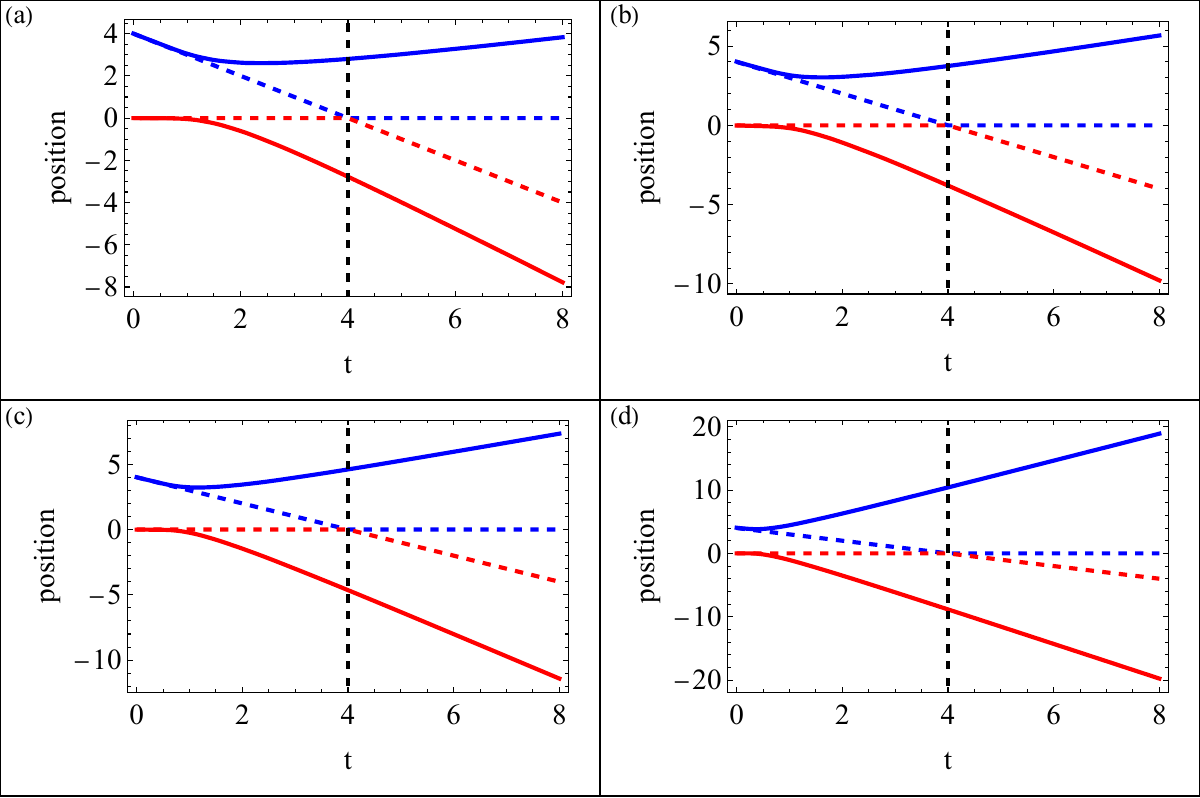}
\caption{
Trajectories of two distinguishable equal-mass particles with a hard-core interaction. Panels (a)-(d) compare the Bohmian ensemble averages, shown by solid lines, with the corresponding classical elastic-collision trajectories, shown by dashed lines. The initial Gaussian widths are $\sigma=0.15,0.20,0.25$, and $0.50$, while $m\sigma^2=0.0625$. The velocities are fixed at $v_1=-1$ and $v_2=0$; therefore $m=0.0625/\sigma^2$ and $p_n=mv_n$ vary from panel to panel. The initial positions are $x_{1,0}=4$, and $x_{2,0}=0$. Blue and red curves denote particles 1 and 2, respectively. The vertical black dashed line marks the common classical collision time $t_c=4$. Each Bohmian average is computed from 300 realizations.
}
\label{fig:hard_trajectories}
\end{figure}

Figure~\ref{fig:hard_trajectories} compares the Bohmian ensemble averages with the classical equal-mass
hard-core trajectories, keeping the fixed scaling $m\sigma^2$. The initial conditions are chosen such that particle 1 moves toward particle 2, which is initially at rest: $x_{1,0} > x_{2,0}, p_1 < 0$, and $p_2 = 0$. In this scan, the initial velocities are fixed at $v_n=p_n/m$ while $m\sigma^2=0.0625$. Consequently, $m=0.0625/\sigma^2, p_n=m(\sigma)v_n$ and the classical collision time $t_c =(x_{1,0} - x_{2,0})/(v_2 - v_1)=4$ is the same in all panels. We have checked that the Bohmian ensemble average and the Born expectation value (not shown) represent the same position mean, cf. Eq. \eqref{eq:meanx}, up to the sampling error. %To separate finite-width effects from the simultaneous mass variation in the fixed-$m\sigma^2$ scan, Figure \ref{fig:s1} in Appendix repeats the trajectory calculation at fixed $m=1$ and initial velocities. The same early-reflection trend persists, confirming that it is not solely caused by the changing mass. 
For packets with nonvanishing widths, the quantum mean can deviate from the point-particle collision curve before the classical collision time because the relative-coordinate wave packet has nonzero support near the hard-core boundary before its center reaches $r = 0$. This early deflection is therefore a broad-width reflection effect induced by the boundary, not a failure of the hard-core model or of the classical limit.

\begin{figure}[H]
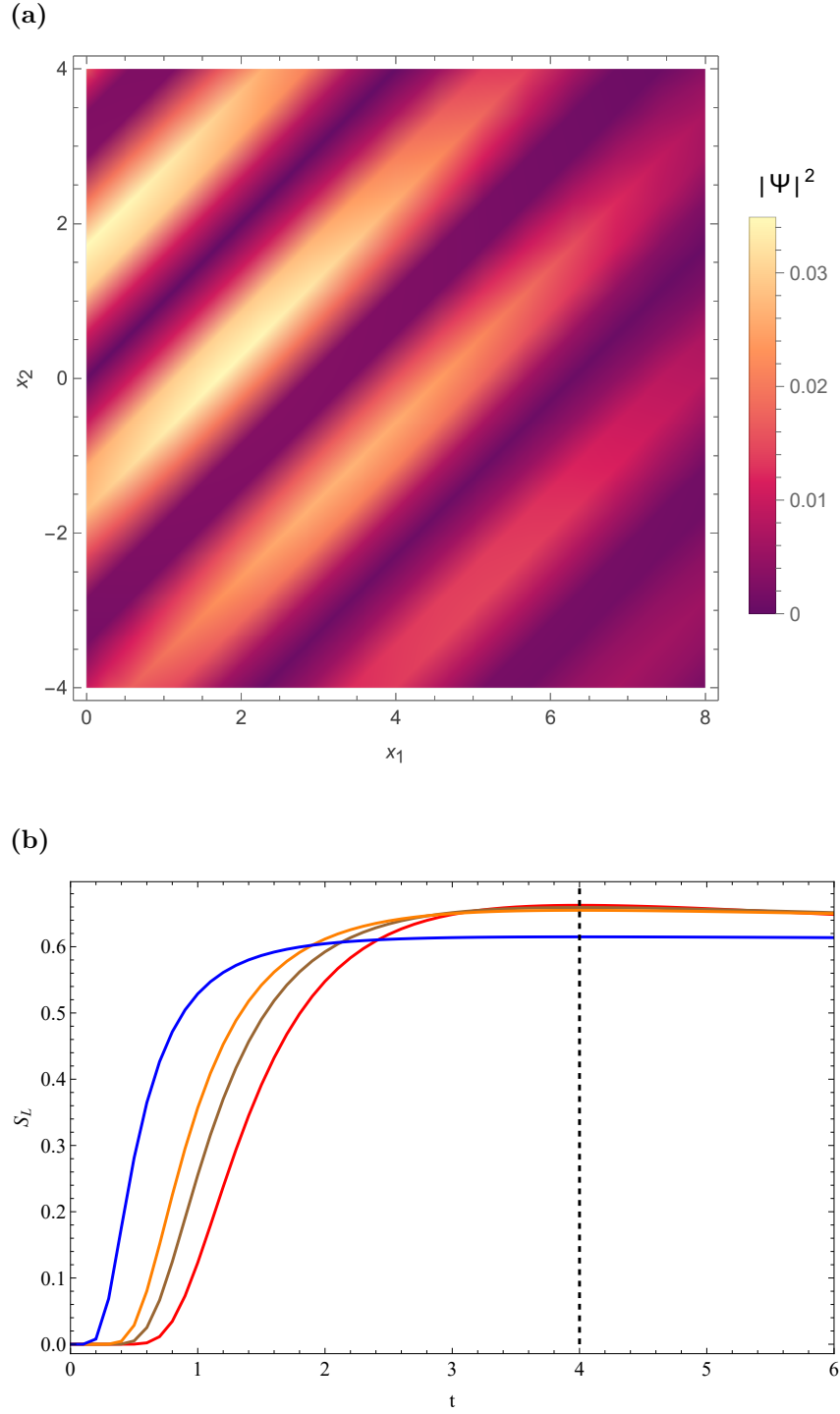

\centering
\subfloat[]{%
\safeincludegraphics[width=0.68\textwidth]{fig2a.pdf}
}
\\
\subfloat[]{%
\safeincludegraphics[width=0.68\textwidth]{fig2b.pdf}
}
\caption{Two-particle probability density and the linear entropy for the hard-core interaction. Panel (a) shows a snapshot of $|\Psi(x_1,x_2,t)|^2$ at $t=2$ for $\sigma=0.25,m=1$. The remaining parameters are those of in Fig.~\ref{fig:hard_trajectories}. Panel (b) shows the linear entropy, cf. Eq. \eqref{eq:linear_entropy_app}, for the hard-core collision, corresponding to the trajectory calculations in Fig.~\ref{fig:hard_trajectories}. Red, brown, orange and blue curves correspond to $\sigma=0.15,0.20,0.25$ and $0.5$, respectively, with $m\sigma^2=0.0625$ fixed. The same fixed velocities, $v_1=-1,v_2=0$, and common collision time $t_c=4$ (marked by the vertical black dashed line) are used as in Fig. \ref{fig:hard_trajectories}.}
\label{fig:hard_entanglement}
\end{figure}

Figure~\ref{fig:hard_entanglement}(a) shows a snapshot of the corresponding two-particle probability distribution $ |\Psi|^2 $ after the boundary reflection. Although the initial state is approximately separable in $x_1$ and $x_2$ because the two packets are well separated, the hard-core boundary subsequently produces a correlated
two-particle density. The interference pattern between the incident and reflected relative-coordinate components is visible once the relative-coordinate packet overlaps the boundary region. After tracing out particle 2, the exact reduced dynamics of particle 1 is described by the reduced density matrix, or equivalently by the marginal density and marginal current. It is not, in general, generated by a closed one-particle Schr\"{o}dinger equation with a local scalar potential. See more discussion in Appendix \ref{app:hard_core_derivation}. Figure \ref{fig:hard_entanglement}(b) shows the corresponding linear entropy. For all parameter sets, $S_L(t)$ begins to increase before the common classical collision time $t_c=4$. This early growth is controlled by the distance from the relative-coordinate packet to the boundary measured in units of its instantaneous width,
\begin{equation}
    \eta(t)=\frac{r_c(t)}{\sigma_r(t)}.
\end{equation}
Larger initial widths give smaller $\eta$ and therefore cause the leading tail of the packet to reach the boundary eariler, consistent with the earlier onset of $S_L$ for broader packets. The late-time magnitude of the entropy, however, is not monotonic in $\sigma$: in the present calculations the broadest packet reaches its plateau earliest but has a slightly smaller plateau value. This behavior shows that the onset of entanglement is governed mainly by boundary overlap, whereas the eventual entropy may depend on the complete Schmidt-mode structure of the scattered state.

The main scan changes $m$ and $p_1$ together with $\sigma$ in order to keep $m\sigma^2$ and the initial velocities fixed. To distinguish this scaling effect from a pure width dependence, Figs. \ref{fig:s1} and \ref{fig:s2} in Appendix repeat the trajectory and entropy calculations at fixed $m=1,v_1=-1$, and $v_2=0$. The early-reflection and entanglement-generation mechanisms persist in this control calculation, showing that they are not artifacts of the mass variation.

% ============================================================
\subsection{Rabi model: covariance-induced breakdown of Ehrenfest closure}
\label{subsec:rabi}
% ============================================================
The performance of Ehrenfest dynamics in two-level systems coupled to bosonic coordinates has been examined extensively, particularly in spin-boson benchmarks and, more recently, in the quantum Rabi model. The Hamiltonian is
\begin{equation}
    H_{\rm Rabi}
    =
    \frac{p^2}{2m}
    +
    \frac{1}{2}m\omega^2x^2
    +
    \frac{\hbar\Omega}{2}\sigma_z
    +
    g_Rx\sigma_x,
    \label{eq:rabi_hamiltonian}
\end{equation}
where $m$ is the oscillator mass, $\omega$ the frequency, $\hbar\Omega$ the two-level splitting, $g_R$ the coupling strength, and $\sigma_{x,y,z}$ refer to Pauli matrices. After expressing the coupling in terms of creation, annihilation, and spin raising/lowering operators, applying the rotating-wave approximation removes the counter-rotating terms and yields the (analytically solvable) Jaynes-Cummings model \cite{Jaynes1963}.

Previous studies have established that mean-field dynamics may fail because it cannot represent decoherence, nuclear or field-mode branching, detailed balance, or quantum-classical correlations \cite{Manfredi2023}. Recent Rabi-model work has also analyzed the semiclassical Hamiltonian limit and the dependence of mixed quantum-classical Rabi oscillations on the initial sampling of the classical mode \cite{TwyeffortIrish2022,Coleman2026,Hsieh2025}. Our narrower purpose is to expose the nonclosure directly in the exact moment equations. As will be shown below, for the Hamiltonian in Eq. \eqref{eq:rabi_hamiltonian}, the oscillator first-moment equations close on $\langle \sigma_x\rangle$, whereas the spin equations require the mixed moments $\langle x\sigma_y\rangle$ and 
$\langle x\sigma_z\rangle$. Their connected parts therefore provide explicit instantaneous source terms omitted by the Ehrenfest factorization.

A mixed quantum-classical Ehrenfest approximation is obtained by replacing the coordinate and momentum operators with classical variables $x(t)$ and $p(t)$ (for the sake of simplicity, we do not distinguish observables and their operators), while retaining a quantum spin state
\begin{equation}
    \ket{\chi(t)}=c_u(t)\ket{\uparrow}+c_d(t)\ket{\downarrow}.
    \label{eq:rabi_spin_state}
\end{equation}
The classical equations are
\begin{align}
    \frac{d  x}{d  t}
    &=
    \frac{p}{m},
    \label{eq:rabi_ehrenfest_x}
    \\
    \frac{d  p}{d  t}
    &=
    -m\omega^2x
    -
    g_R
    \bra{\chi(t)}\sigma_x\ket{\chi(t)},
    \label{eq:rabi_ehrenfest_p}
\end{align}
whereas the spin evolves according to
\begin{equation}
    i\hbar
    \frac{d }{d  t}
    \begin{pmatrix}
    c_u\\
    c_d
    \end{pmatrix}
    =
    \begin{pmatrix}
    \hbar\Omega/2 & g_Rx\\
    g_Rx & -\hbar\Omega/2
    \end{pmatrix}
    \begin{pmatrix}
    c_u\\
    c_d
    \end{pmatrix}.
    \label{eq:rabi_spin_equation}
\end{equation}

The origin of the Ehrenfest error can be stated more explicitly. In the full quantum model, the mean oscillator variables obey
\begin{align}
    \frac{d \expect{x}}{d  t}
    &=
    \frac{\expect{p}}{m},
    \\
    \frac{d \expect{p}}{d  t}
    &=
    -m\omega^2\expect{x}
    -
    g_R\expect{\sigma_x},
\end{align}
and the spin expectation values satisfy
\begin{subequations}
\begin{align}
    \frac{d \expect{\sigma_x}}{d  t}
    &=
    -\Omega\expect{\sigma_y},
    \\
    \frac{d \expect{\sigma_y}}{d  t}
    &=
    \Omega\expect{\sigma_x}
    -
    \frac{2g_R}{\hbar}\expect{x\sigma_z},\label{eq:rabi_exact_spin_momentsb}
    \\
    \frac{d \expect{\sigma_z}}{d  t}
    &=
    \frac{2g_R}{\hbar}
    \expect{x\sigma_y}.
    \label{eq:rabi_exact_spin_momentsc}
\end{align}
\end{subequations}
Unlike the generic smooth-potential correction in Eq. (17), the nonclosure in the Rabi model does not originate from curvature of the oscillator force. The oscillator mean equation is exact at the first-moment level. The hierarchy becomes nonclosed through the spin equations, which involve the mixed moments $\expect{x\sigma_y}$ and $\expect{x\sigma_z}$.
%The mixed quantum-classical Ehrenfest approximation replaces the mixed moments by factorized products,
%\begin{equation}
%    \expect{x\sigma_\alpha}
%    \approx
%    \expect{x}\expect{\sigma_\alpha},
%    \qquad
%    \alpha=y,z.
%\end{equation}
%Therefore, the breakdown of the semiclassical closure is controlled by the spin-position covariances
%\begin{equation}
%    C_{x\sigma_\alpha}
%    =
%    \expect{x\sigma_\alpha}
%    -
%    \expect{x}\expect{\sigma_\alpha},
%    \qquad
%    \alpha=y,z.
%    \label{eq:rabi_covariances}
%\end{equation}
A change in the initial wave-packet width affects the semiclassical error only indirectly, by changing the magnitude and persistence of these covariances. In the numerical results below, this mechanism is tested by comparing the full quantum dynamics with the Ehrenfest approximation and by monitoring the growth of the spin-position covariances.

\begin{figure}[H]
\centering
\safeincludegraphics[width=0.8\textwidth]{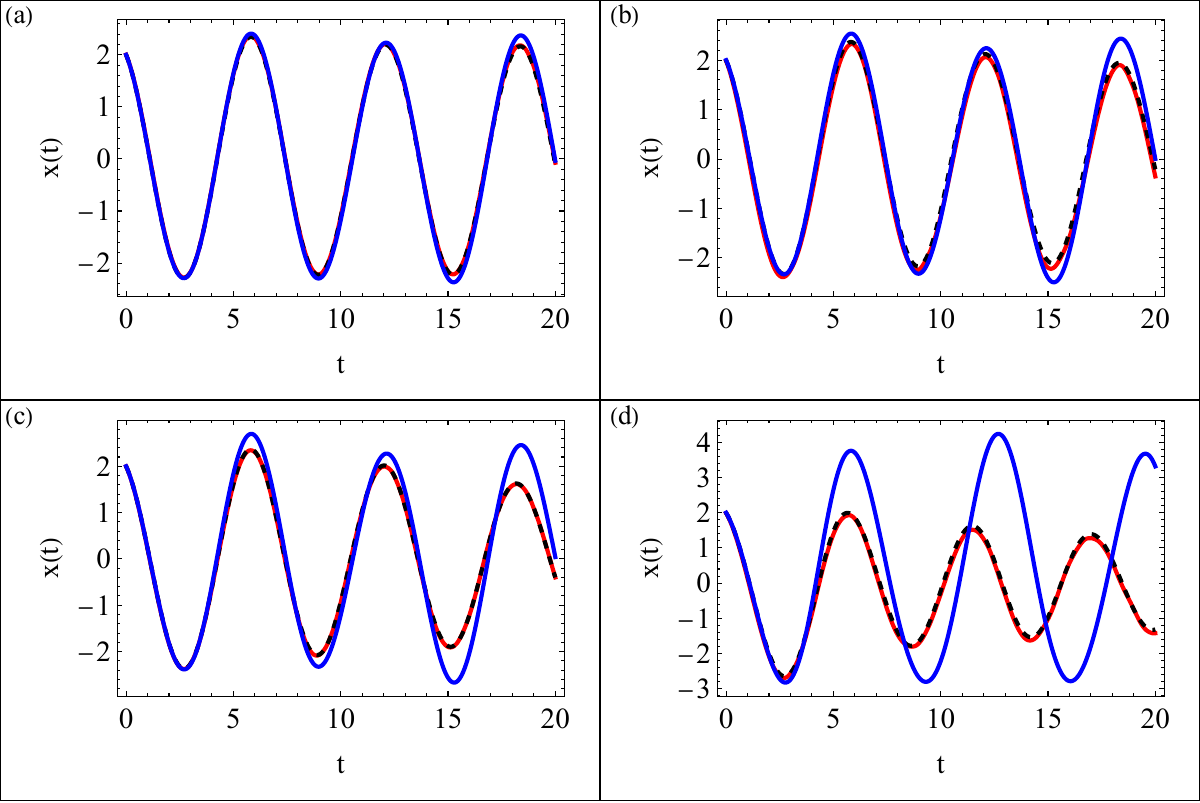}
\caption{
Oscillator trajectories in the Rabi model. Panels (a)--(d) compare the average Bohmian trajectory (red solid lines), the Born expectation value (black dashed lines), and the mixed quantum-classical Ehrenfest trajectory (blue solid lines) for initial wave-packet widths $\sigma=0.15,0.20,0.25$, and $0.50$, with $m\sigma^2=0.0625$ fixed. The initial conditions are $x_0=2$ and $v_0=-1$ ($p_0=mv_0$ varies)  and the up state $|\uparrow\rangle$ is populated. The Bohmian average is obtained from 300 realizations. We use units with $\hbar=\omega=1$. Other parameters are $\Omega=1$ and $g_R=0.25$. }
\label{fig:rabi_trajectory}
\end{figure}

\begin{figure}[H]
\centering
\safeincludegraphics[width=0.8\textwidth]{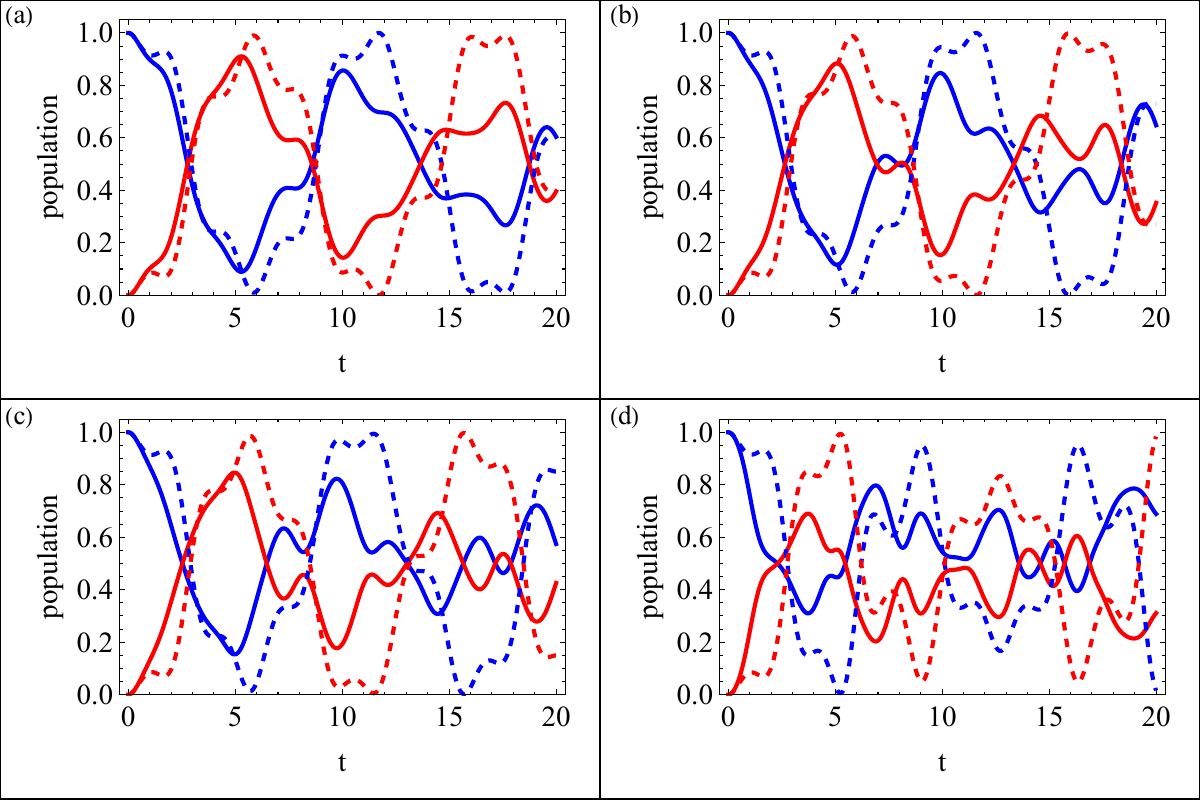}
\caption{
Spin populations in the Rabi model for the same parameters as in Fig.~\ref{fig:rabi_trajectory}. Blue and red curves denote the two spin components $|\uparrow\rangle, |\downarrow\rangle$, respectively. Solid curves are obtained from the full quantum calculation (Born expectation), while dashed curves correspond to the mixed quantum-classical Ehrenfest approximation.}
\label{fig:rabi_population}
\end{figure}

Figures~\ref{fig:rabi_trajectory} and \ref{fig:rabi_population} compare the full quantum dynamics with the mixed quantum-classical Ehrenfest approximation along the fixed $m\sigma^2$, fixed initial velocity sequence. The oscillator is initialized in a Gaussian wave packet associated with a definite spin component ($\chi=|\uparrow\rangle$ at $t=0$). The Bohmian sample mean agrees with the Born expectation value within finite-sampling error, as required by equivariance. The quantum and Ehrenfest results agree most closely for the narrowest members of this sequence, while larger deviations develop for the broader, lighter packets at later times. This trend should not be interpreted as a pure dependence on the initial width, because increasing $\sigma$ simultaneously decreases the oscillator mass according to $m=\kappa/\sigma^2$, while the initial momentum changes as $p_0=mv_0$. The origin of the Ehrenfest error in spin dynamics is identified by the mixed moments in Eqs. \eqref{eq:rabi_exact_spin_momentsb} and \eqref{eq:rabi_exact_spin_momentsc}. Using
\begin{equation}
    \expect{x\sigma_\alpha}=\expect{x}\expect{\sigma_\alpha}+C_{x\sigma_\alpha},\quad \alpha=y,z,
\end{equation}
shows that the Ehrenfest factorization omits the instantaneous terms
\begin{equation}
    -\frac{2g_R}{\hbar}C_{x\sigma_z}\quad\text{and}\quad \frac{2g_R}{\hbar}C_{x\sigma_y}
\end{equation}
from the equations for $\expect{\sigma_y}$ and $\expect{\sigma_z}$, respectively. We therefore define
\begin{equation}
    \xi_{\text{cov}}=\frac{2g_R}{\hbar\omega}\sqrt{|C_{x\sigma_y}(t)|^2+|C_{x\sigma_z}(t)|^2},
    \label{eq:spin_position_covariance}
\end{equation}
as a dimensionless magnitude of the covariance terms omitted by the Ehrenfest closure, which vanishes when the mixed moments factorize exactly. The influence of the covariances on the position mean is indirect: they first modify the spin dynamics, which then feeds back into the oscillator force through $\langle \sigma_x\rangle$.

\begin{figure}[H]
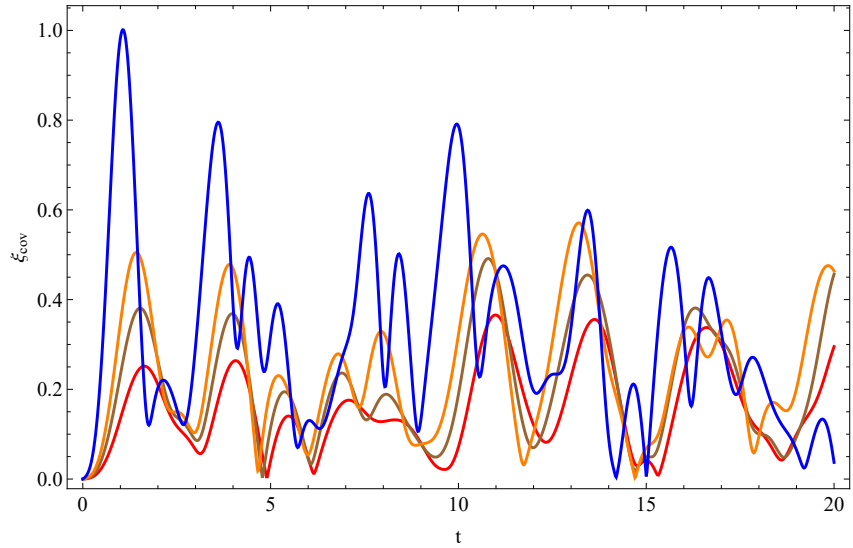

\centering
\safeincludegraphics[width=0.68\textwidth]{fig5.pdf}
\caption{Dimensionless covariance-induced closure-error measure, cf. Eq. \eqref{eq:spin_position_covariance}, for the Rabi-model calculations in Figs.~\ref{fig:rabi_trajectory} and \ref{fig:rabi_population}. Red, brown, orange and blue lines correspond to $\sigma=0.15,0.2,0.25$ and $0.5$, along the fixed-$m\sigma^2$, fixed-initial-velocity sequence.}
\label{fig:rabi_covariance}
\end{figure}

Figure \ref{fig:rabi_covariance} shows that $\xi_{cov}$ is appreciable throughout the evolution and varies systematically along the fixed-$m\sigma^2$, fixed initial velocity scaling as $\sigma$ increases. The broader members of this sequence exhibit a more persistent late-time closure error, consistent with the increasing quantum–Ehrenfest deviations in Figs. \ref{fig:rabi_trajectory} and \ref{fig:rabi_population}. However, this trend combines changes in the packet width, oscillator mass, and initial momentum. It should therefore be interpreted as a property of the complete scaling sequence rather than as evidence for a universal monotonic dependence on $\sigma$.

Fig. \ref{fig:s3} in Appendix provides the complementary fixed-mass, fixed-velocity control. In that calculation, only the initial width is varied. The resulting $\xi_{cov}(t)$ is nonmonotonic in $\sigma$: narrower packets can produce larger and sharper early-time peaks, whereas broader packets can exhibit more persistent late-time correlations and a larger accumulated displacement error. This confirms that the packet width controls not only the magnitude but also the temporal structure of the spin-position correlations. The covariance measure is an instantaneous source term in the nonclosed spin equations; it is not itself equal to the accumulated position difference $\langle x(t)\rangle-x_{semi}(t)$. The latter develops through the coupled oscillator-spin dynamics and therefore need not peak at the same times, or follow the same ordering in $\sigma$.

Differences between the full quantum and semiclassical spin populations in Fig. 4 consequently reflect the fact that the spin evolves under the spatially distributed quantum state of the oscillator rather than solely under its mean coordinate. The coupling-strength scan in Fig. \ref{fig:s4} provides a separate control by fixing $m$ and $\sigma$ while varying $g_R$. We also note that the local spinor weights evaluated along a single Bohmian trajectory need not remain constant; an illustrative example is provided in Fig. \ref{fig:s5} in the Appendix.

%The coupling-strength scan in Fig. \ref{fig:s4} in Appendix isolates a different control parameter by fixing $m=1$ and $\sigma=0.25$ while varying $g_R$. The covariance measure is an instantaneous closure-error scale, not the accumulated displacement error itself. The position difference $\expect{x}-x_{semi}(t)$ develops through coupled oscillator-spin dynamics and therefore need not peak at the same time as $\xi_{cov}(t)$, nor vary monotonically with either raw covariance component. Nevertheless, Fig. \ref{fig:s4} shows that increasing $g_R$ amplifies the covariance contribution and is accompanied by a larger quantum–Ehrenfest position discrepancy. Further, Fig. \ref{fig:s5} in Appendix plots the instantaneous spin probability along a representative Bohmian trajectory, showing that individual trajectory ``spin" variables need not stay fixed even if ensemble populations are approximately constant.

% ============================================================
\subsection{Internal-state-dependent force: exact mean but branch-resolved pointer separation}
\label{subsec:internal_state}
% ============================================================
In the last example, we consider an ISDF model involving a free particle coupled to a two-level system $\{\ket{g},\ket{e}\}$:
\begin{equation}
    H_{\rm IS}
    =
    \frac{p^2}{2m}
    +
    (E_g+fx)\ket{g}\bra{g}
    +
    (E_e-fx)\ket{e}\bra{e}.
    \label{eq:internal_hamiltonian}
\end{equation}
With the sign convention in Eq.~\eqref{eq:internal_hamiltonian}, the $\ket{g}$ component experiences the force $-f$ along the $-x$ direction, while the $\ket{e}$ component receives the force $+f$ aligned with the $+x$ axis. The total wavefunction at $t=0$ takes the form
\begin{equation}
    \ket{\Psi(0)}
    =
    \psi_0(x)
    \left(
    c_g\ket{g}
    +
    c_e\ket{e}
    \right),
    \qquad
    |c_g|^2+|c_e|^2=1.
    \label{eq:internal_initial_state}
\end{equation}
The spatial part is a Gaussian wave packet carrying zero (mean) momentum,
\begin{equation}
    \psi_0(x)
    =
    \frac{1}{(2\pi\sigma^2)^{1/4}}
    \exp\left[
    -\frac{(x-x_{0})^2}{4\sigma^2}
    \right].
    \label{eq:initial_gaussians}
\end{equation}
There is no direct coupling between $\ket{g}$ and $\ket{e}$, so the internal-state populations are conserved. That is, the populations of internal states
\begin{equation}
    P_g=|c_g|^2,
    \qquad
    P_e=|c_e|^2
    \label{eq:internal_populations}
\end{equation}
remain constant. The full wavefunction at later times can be written as
\begin{equation}
    |\Psi(t)\rangle
    =
    c_g\psi_{-f}(x,t)|g\rangle
    +
    c_e\psi_{+f}(x,t)|e\rangle,
    \label{eq:internal_time_state}
\end{equation}
where $\psi_F(x,t)$ denotes the spatial wavefunction propagated under the constant force $F$:
\begin{equation}
    \psi_F(x,t)
    =
    e^{-i E_Ft/\hbar}
    \exp\left[
    \frac{i}{\hbar}
    \left(
    Ftx-\frac{F^2t^3}{6m}
    \right)
    \right]
    \phi_{\rm free}
    \left(
    x-\frac{Ft^2}{2m},t
    \right).
    \label{eq:constant_force_solution}
\end{equation}
Here $F=-f$ and $E_F=E_g$ for the $|g\rangle$ branch, while $F=+f$ and $E_F=E_e$ for the $|e\rangle$ branch.

The total position density is
\begin{equation}
    \rho(x,t)
    =
    P_g n_g(x,t)
    +
    P_e n_e(x,t),
    \label{eq:probdensity}
\end{equation}
with
\begin{equation}
    n_g(x,t)
    =
    \frac{1}{\sqrt{2\pi}\sigma(t)}
    \exp\left[
    -\frac{(x-x_g(t))^2}{2\sigma(t)^2}
    \right],
\end{equation}
\begin{equation}
    n_e(x,t)
    =
    \frac{1}{\sqrt{2\pi}\sigma(t)}
    \exp\left[
    -\frac{(x-x_e(t))^2}{2\sigma(t)^2}
    \right],
\end{equation}
where the packet width is defined in Eq. \eqref{eq:sigma_t}.
The branch centers are
\begin{equation}
    x_g(t)=x_0-\frac{ft^2}{2m},
    \qquad
    x_e(t)=x_0+\frac{ft^2}{2m}.
    \label{eq:internal_branch_centers}
\end{equation}
The Born mean position is therefore
\begin{equation}
    \expect{x(t)}
    =
    x_0
    +
    \frac{f(P_e-P_g)}{2m}t^2.
    \label{eq:internal_exact_mean}
\end{equation}

\begin{figure}[H]
\centering
\safeincludegraphics[width=0.8\textwidth]{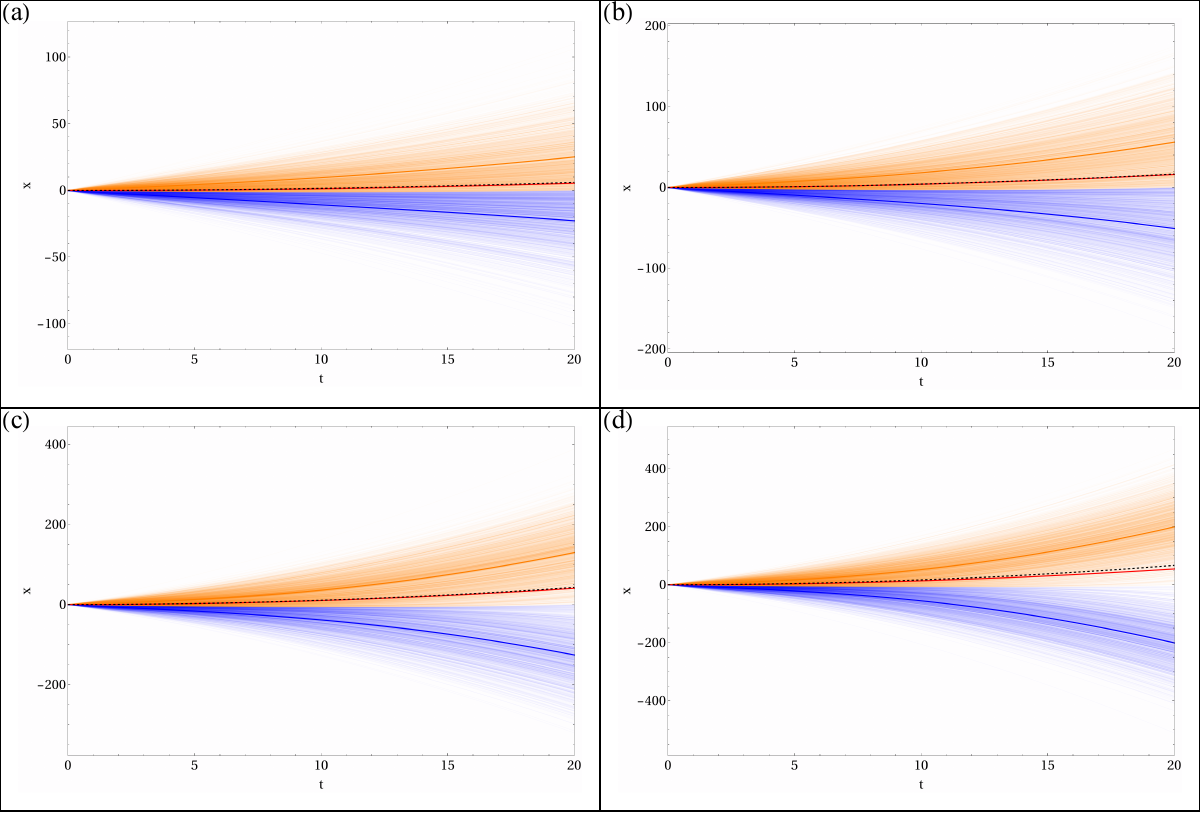}
\caption{Bohmian trajectories for the ISDF model in Eq.~\eqref{eq:internal_hamiltonian}, compared with the exact analytical mean. Panels (a)--(d) correspond to Gaussian wave packets with initial widths $\sigma=0.15,0.25,0.40$, and $0.50$, while $m\sigma^2$ is fixed. The red solid curves show the means of 3000 Bohmian trajectories, while the black dashed curves show the semiclassical Ehrenfest trajectory, cf. Eq.~\eqref{eq:internal_exact_mean}. Their agreement is consistent with the finite-sample diagnostics. The faint yellow and faint blue trajectories are postselected according to whether their final positions (at $t_{max}=20$) are positive or negative; the solid yellow and blue curves show the corresponding conditional averages. Energies are $E_g=0$ and $E_e=1$, the initial position of the packet peak is $x_0=0$, the force parameter is $f=0.25$ and initial internal-state populations are $P_g=1/3$ and $P_e=2/3$.}
\label{fig:internal_preliminary}
\end{figure}

The spinor Bohmian velocity can be written explicitly as
\begin{equation}
    v(x,t)
    =
    \frac{
    P_g n_g(x,t)v_g(x,t)
    +
    P_e n_e(x,t)v_e(x,t)
    }{
    P_g n_g(x,t)+P_e n_e(x,t)
    },
    \label{eq:internal_bohmian_velocity}
\end{equation}
where
\begin{subequations}
\begin{equation}
    v_g(x,t)
    =
    -\frac{ft}{m}
    +
    A(t)\left[x-x_g(t)\right],
\end{equation}
\begin{equation}
    v_e(x,t)
    =
    +\frac{ft}{m}
    +
    A(t)\left[x-x_e(t)\right],
\end{equation}
and
\begin{equation}
    A(t)
    =
    \frac{\hbar^2t}{4m^2\sigma^2\sigma(t)^2}.
\end{equation}
\end{subequations}
Figure~\ref{fig:internal_preliminary} shows Bohmian trajectory ensembles, ensemble averages, and Born expectation values for Gaussian wave packets with different initial widths, while keeping $m\sigma^2$ fixed. We have verified that the corresponding internal-state populations remain exactly constant (not shown). Because the potentials in Eq.~\eqref{eq:internal_hamiltonian} are linear in $x$, the Ehrenfest equation for the total mean position closes exactly. Therefore, the Born expectation value in Eq.~\eqref{eq:internal_exact_mean} coincides with the semiclassical Ehrenfest prediction. %In the infinite-sampling limit, it also coincides with the average Bohmian trajectory.

As the initial width $\sigma$ decreases, the Bohmian ensemble becomes increasingly concentrated around the classical counterpart of Eq.~\eqref{eq:internal_exact_mean}. For broader packets, the ensemble samples a wider range of positions. The relevant quantum effect in this model is not a deviation of the total mean position from Ehrenfest dynamics, but the branch-resolved separation of the two internal components. The two branches associated with $\ket{g}$ and $\ket{e}$ have centers $x_g(t)$ and $x_e(t)$, given in Eq.~\eqref{eq:internal_branch_centers}. Their separation is
\begin{equation}
    \Delta x(t)
    =
    x_e(t)-x_g(t)
    =
    \frac{ft^2}{m}.
    \label{eq:branch_separation}
\end{equation}

Since $\ket{g}$ and $\ket{e}$ are orthogonal, there is no interference term in the total position density. The total density is the incoherent sum of the two branch-resolved Gaussian densities, as shown in Eq.~\eqref{eq:probdensity}. Nevertheless, when the two branch densities overlap, the branch centers are not equal to the conditional averages of trajectories located in the $+x$ or $-x$ regions. This distinction is essential for obtaining the correct Born expectation value and the correct Bohmian velocity in Eq.~\eqref{eq:internal_bohmian_velocity}.

\begin{figure}[H]
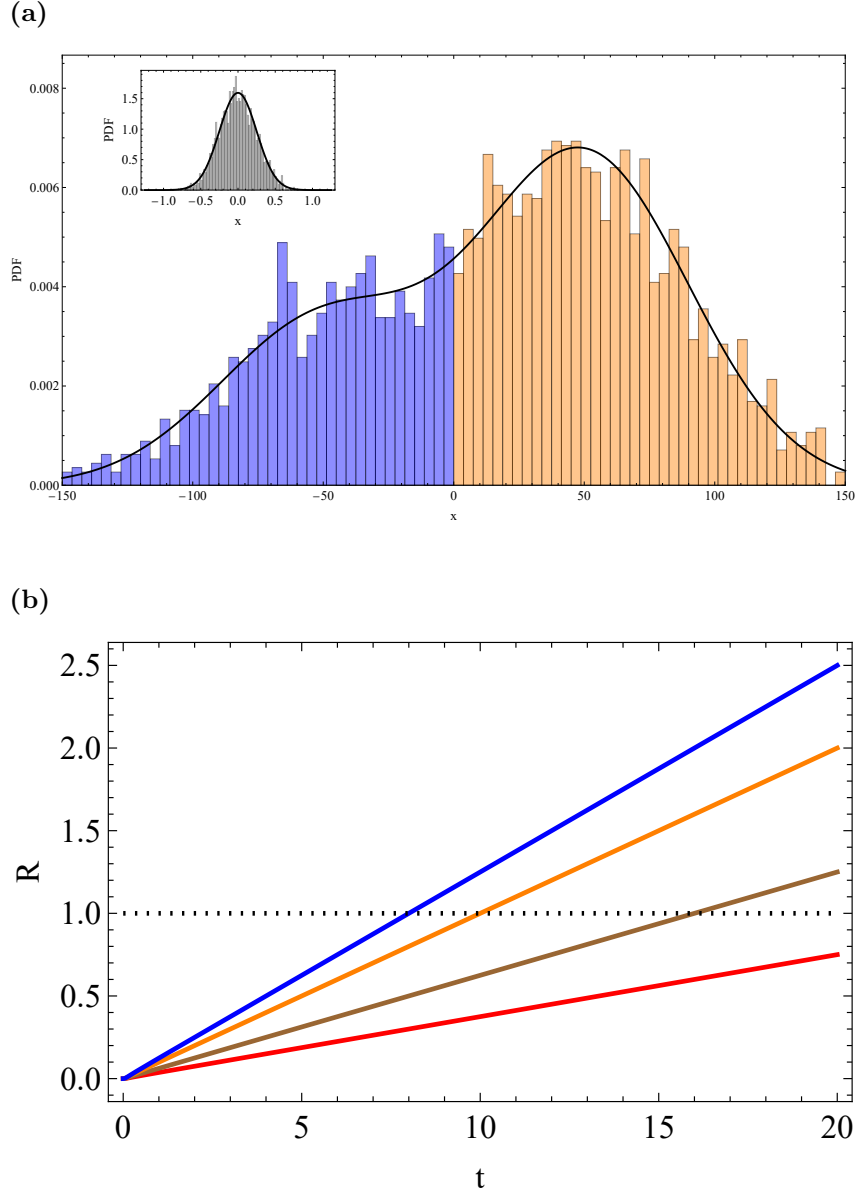

\centering
\subfloat[]{%
\safeincludegraphics[width=0.68\textwidth]{fig7a.pdf}
}
\\
\subfloat[]{%
\safeincludegraphics[width=0.68\textwidth]{fig7b.pdf}
}
\caption{Positions statistics and branch resolution in the ISDF model. Panel (a) shows the histogram of 3000 Bohmian trajectories at $t=20$; the inset showing the initial distribution. The black solid curve is analytical total density in Eq.~\eqref{eq:probdensity} for $\sigma=0.25$ and $m=1$. Panel (b) shows the branch-resolution parameter $R(t)$ defined in Eq. \eqref{eq:branch_resolution}. Red, brown, orange and blue lines correspond to $\sigma=0.15,0.25,0.4$ and $0.5$, respectively, along the fixed-$m\sigma^2$ sequence. The horizontal dotted line at $R=1$ marks a heuristic crossover between strongly overlapping and increasingly resolved branches; it is not a sharp threshold. Other parameters are the same as Fig. \ref{fig:internal_preliminary}.}
\label{fig:internal_statistics}
\end{figure}

For the internal-state configuration $P_g=1/3$ and $P_e=2/3$, the total mean force is positive, and the mean trajectory shifts toward the $+x$ region. Figure~\ref{fig:internal_statistics}(a) compares the Bohmian sample statistics with the analytical probability density. Note that finite-sampling effects, boundary reflections, and resolution errors do not bring physically significant changes to the displayed results. We emphasize that individual Bohmian trajectories do not literally split into two copies. Each trajectory follows a smooth path. The word ``branching'' refers to the ensemble-level separation of trajectories guided by the two spatial lobes of the spinor wavefunction. At finite times, especially when the two branch densities overlap, the ratio of the number of trajectories found in the $+x$ and $-x$ regions need not coincide with $P_e/P_g$. Only when the branches are well separated and their overlap is negligible can spatial regions be approximately associated with the corresponding internal-state branches. % For fixed $m\sigma^2$, decreasing $\sigma$ reduces the initial spatial spread but changes the mass according to the chosen scaling.
The branch-resolution parameter $R(t)$ defined in Eq. \eqref{eq:branch_resolution} provides a more explicit spatial branch resolution measurement: $R\sim1$ is used as a heuristic crossover between strongly overlapping and increasingly resolved branches; it is not a sharp dynamical or measurement threshold. In Fig. \ref{fig:internal_statistics}(b), we compare graphs of $R(t)$ for different choices of $\sigma$, where broader (initial) widths lead to larger $R(t)$ at each snapshot, which itself increases in time, indicating the process of the branch separation. The overlap of the two position densities can be quantified by the Bhattacharyya coefficient,
\begin{equation}
    \mathcal{B}_\rho(t)
    =
    \int_{-\infty}^{\infty}
    \sqrt{n_g(x,t)n_e(x,t)}
    d x
    =
    \exp\left[-\frac{R^2(t)}{2}\right].
\end{equation}
This overlap approaches 1 when  $R\ll1$ (branches completely overlap) and decays to 0 when $R\gg1$ (branches are well separated). %This diagnostic measures when the total mean position can be Ehrenfest-like even if the trajectory ensemble separates into branch-resolved groups.

Figures~\ref{fig:s6} and \ref{fig:s7} in Appendix show additional trajectory statistics for different internal-state populations. As long as the packet has a nonzero width, even when the internal state is purely $\ket{e}$, some trajectories can still fall in the $-x$ region within the chosen time window because the spatial wave packet spreads in time. In contrast, for the equal coherent superposition $P_g=P_e=1/2$, the total mean position remains on the $x=0$ axis, while the ensemble separates into two approximately symmetric groups at all times.

% ============================================================
\section{Conclusion}
\label{sec:conclusion}
% ============================================================
We have analyzed the quantum-to-classical transition of interacting systems from the viewpoint of reduced dynamical closure. While some principal closure can be formulated without Bohmian ontology, the trajectory representation supplies a realization-resolved probability-flow diagnostic that permits direct comparison among individual realizations, Born-rule ensemble averages, and classical or mixed quantum-classical equations of motion.

The hard-core collision demonstrates how distributional classicality can fail even when mean-level classicality is maintained: even a thin (nonzero) relative-coordinate packet has support near the contact boundary before its center arrives, reflection and interference are prior to the classical collision and generate spatial nonseparability, producing spatial entanglement (failure of distributional classicality) despite the average motion remaining close to classical. The exact one-coordinate dynamics is specified by the reduced density and marginal current, rather than by a closed local scalar potential.

The Rabi model exhibits a different mechanism. The oscillator mean equation is exact at the first-moment level, but the spin equations involve the mixed moments $\expect{x\sigma_y}$ and $\expect{x\sigma_z}$. Therefore, even when an oscillator’s mean coordinate follows a classical-like trajectory (mean-level classicality), spin-position covariances can prevent full semiclassical closure. The fixed-mass control further shows that the dependence of this nonclosure on the initial width is dynamical and nonmonotonic: width changes the temporal profile of the covariance source rather than simply setting its magnitude. In contrast, the ISDF model shows that the total position mean obeys an exact Ehrenfest equation because the branch forces are linear and the populations are conserved. Nevertheless, the state develops branch-resolved pointer structure. Hence, an exact Ehrenfest (mean-classical) coordinate can hide underlying branch formation (lack of branch-level resolution). %Mean-level classicality therefore does not imply distributional or branch-level classicality.

These results show that for localized Gaussian wave packets, the Bohmian ensemble becomes increasingly concentrated around the classical trajectory as the packet width is reduced with an appropriate mass-width scaling. By virtue of quantum equilibrium, the Bohmian ensemble mean converges to the quantum expectation value in the limit of infinite trajectories. When the wave packet is sufficiently narrow, this common quantum mean itself also follows classical trajectory. A concurrent analysis also argues that Ehrenfest dynamics is exact in the strict delta-localized classical limit and explicitly excludes trajectory bifurcation from that limit \cite{Nazarov2026}. This is consistent with our findings: the localized scaling sequence approaches Ehrenfest behavior.

For wide wave packets, however, the full quantum dynamics usually develop correlations (e.g., interference, covariance and/or branch structure) that obstruct a closed single-trajectory description. Thus, the emergence of classical reduced dynamics depends not only on spatial localization but also on how interaction-induced correlations enter the equations for the observables being retained. Future work may extend this analysis to open-system projection methods \cite{Breuer2007,Nakajima1958,Zwanzig1960}, decoherence \cite{vonNeumann1955,Zurek1993}, and many-body dynamics, where memory effects, environmental entanglement, and coarse graining provide additional mechanisms of reduced dynamical nonclosure.

\appendix

\section{Numerical Methods and Convergence Checks}
\label{app:numerical}
In this section, we introduce the numerical methods implemented. All calculations use $\hbar=1$. Random initial configurations are generated with pseudorandom seed \textit{1234}. Bohmian initial positions are sampled from the relevant Born distribution, and trajectories are then propagated from the probability-current velocity. For scalar and two-component wavefunctions, respectively,
\begin{equation}
    v(x,t)
    =
    \frac{\hbar}{m}
    \operatorname{Im}
    \left[
    \frac{\partial_x\psi(x,t)}{\psi(x,t)}
    \right],
\end{equation}
and,
\begin{equation}
    \rho(x,t)
    =
    \sum_{\alpha=1}^{2}
    |\Psi_\alpha(x,t)|^2,
    \quad
    j(x,t)
    =
    \frac{\hbar}{m}
    \operatorname{Im}
    \left[
    \sum_{\alpha=1}^{2}
    \Psi_\alpha^*(x,t)
    \partial_x\Psi_\alpha(x,t)
    \right],\quad v=\frac{j}{\rho}.
\end{equation}
The Bohmian ensemble mean is compared with the Born expectation value as a sampling diagnostic.

\subsection{Hard-core trajectories and linear entropy}

For the hard-core model, the wavefunction is evaluated analytically from the center-of-mass Gaussian and the image-antisymmetrized relative-coordinate wavefunction in Eq. \eqref{eq:hard_core_solution} in the main text. The Bohmian velocities are obtained from logarithmic derivatives of the analytical wavefunction,
\begin{align}
    v_1
    &=
    \frac{\hbar}{m}
    \operatorname{Im}
    \left[
    \frac{1}{2}
    \frac{\partial_X\psi_{\rm CM}}{\psi_{\rm CM}}
    +
    \frac{\partial_r\phi_{\rm HC}}{\phi_{\rm HC}}
    \right],
    \\
    v_2
    &=
    \frac{\hbar}{m}
    \operatorname{Im}
    \left[
    \frac{1}{2}
    \frac{\partial_X\psi_{\rm CM}}{\psi_{\rm CM}}
    -
    \frac{\partial_r\phi_{\rm HC}}{\phi_{\rm HC}}
    \right],
\end{align}
where
\begin{equation}
    \phi_{\rm HC}(r,t)=\phi_{\rm free}(r,t)-\phi_{\rm free}(-r,t).
\end{equation}
In the scan, $\sigma=0.15,0.2,0.25$ and $0.5, m\sigma^2=0.0625$ and the initial velocities are fixed at $v_1=-1$ and $v_2=0$. Thus $m=0.0625/\sigma^2$ and $p_n=mv_n$. For each $\sigma$, $N=300$ ordered initial configurations are generated from independent normal distributions centered at $x_{1,0}=4$ and $x_{2,0}=0$, both with standard deviation $\sigma$; proposals with $x_1\leq x_2$ are rejected. All 300 trajectories were successfully propagated over the complete interval for the displayed calculations. The analytical image wavefunction is used to evaluate the velocity field. The coupled guidance equations are integrated with an adaptive solver to $t_{max}=8$, with maximum step 0.02, accuracy goal 9, and precision goal 9. Trajectories are updated every 0.01 time unit. A numerical safeguard terminates a trajectory if $x_1-x_2<10^{-7}$; no such termination occurred in the displayed calculations. % Failed or incomplete trajectories are excluded.%; the retained number must be reported for each parameter set.

The linear entropy is evaluated every $0.1$ time unit on a time-adaptive $N_x\times N_x$ grid. At each time, the grid follows the approximate reflected particle centers,
\begin{align}
    x_{1,c}&=X_c(t)+\frac{|r_c(t)|}{2};\\
    x_{2,c}&=X_c(t)-\frac{|r_c(t)|}{2},
\end{align}
and covers seven estimated instantaneous one-particle widths on either side, together with an additional margin equal to the initial width $\sigma$. The estimated one-particle width is
\begin{equation}
    \sigma_{1p}(t)=\left[\sigma_X^2(t)+\frac{\sigma_r^2(t)}{4}\right]^{1/2}.
\end{equation}
The main calculations use $N_x=480$ grid points in each coordinate. The forbidden region $x_1\leq x_2$ is set to zero. After normalization of the discretized two-coordinate wavefunction, the reduced density matrix is formed as
\begin{equation}
    \rho_1=B^\dagger B,\quad B_{ij}
    =
    \Delta x\,
    \frac{\Psi(x_i,x_j,t)}
    {
    \left[
    (\Delta x)^2
    \sum_{ij}|\Psi(x_i,x_j,t)|^2
    \right]^{1/2}
    }.
\end{equation}
Its squared singular values define the discrete Schmidt weights $\lambda_v$, with
\begin{equation}
    \text{Tr}\rho_1^2=\sum_{v}\lambda_v^2,
\end{equation}
from which the linear entropy is
\begin{equation}
    S_L(t)=1-\sum_{v}\lambda_v^2.
\end{equation}
In practice, Schmidt weights below $10^{-12}$ were discarded and the remainder were renormalized before the entropy is evaluated. The maximum discarded weight is less than $10^{-8}$ in all cases, ensuring negligible error from this truncation.

\subsection{Rabi-model propagation}

The Rabi wavefunction is represented as a two-component spinor on a uniform spatial grid. Its time evolution is calculated with a second-order split-operator scheme,
\begin{equation}
    \boldsymbol{\Psi}(t+\Delta t)
    \simeq
    e^{-i \hat V\Delta t/(2\hbar)}
    e^{-i \hat T\Delta t/\hbar}
    e^{-i \hat V\Delta t/(2\hbar)}
    \boldsymbol{\Psi}(t),
\end{equation}
where
\begin{equation}
    \hat T
    =
    \frac{\hat p^2}{2m},
\end{equation}
and
\begin{equation}
    \hat V(x)
    =
    \frac{1}{2}m\omega^2\hat{x}^2\mathbbm{1}
    +
    \frac{\hbar\Omega}{2}\hat{\sigma}_z
    +
    g_R\hat{x}\hat{\sigma}_x.
\end{equation}
The kinetic step is evaluated in momentum space using fast Fourier transforms, while the local $2\times2$ potential propagator is evaluated at each spatial grid point. Spatial derivatives entering the Bohmian current are evaluated spectrally.

The spinor wavefunction is propagated on the periodic FFT grid $x\in[-20,20)$ with $N_x=2048$ and $\Delta x=0.01953125$. A second-order split-operator method is used with time step $\Delta t=0.005$, and the wavefunction is stored every 0.02 up to $t_{max}=20$. The kinetic propagator is applied in momentum space, while the local oscillator-plus-spin $2\times 2$ propagator is evaluated exactly at each grid point. Spatial derivatives in the current are evaluated spectrally.

We vary $\sigma$ while holding $m\sigma^2=0.0625$ and the initial velocity $v_0=-1$ (so the initial momentum $p_0=mv_0$ varies with $m$). Other parameters are $x_0=2,g_R=0.25,\hbar=\Omega=\omega=1$. For each $\sigma$, $N=300$ initial positions are sampled from the initial Gaussian density. The velocity field is interpolated cubically in space and time, and trajectories are propagated with a fourth-order Runge–Kutta scheme using step 0.02. The numerical velocity is regularized as $v=j/(\rho+10^{-12})$. The implementation monitors attempted exits from the interpolation domain. Any trajectory leaving the spatial grid (beyond $x=\pm19$) is stopped and excluded from the average to prevent artifacts from periodic boundary conditions. No clipping events occurred in the data reported here. We also verified that the Bohmian ensemble mean agrees with the quantum expectation to within statistical sampling error (deviations $\lesssim10^{-3}$ for 300 trajectories).

The mixed moments are calculated as
\begin{equation}
    \expect{x\sigma_\alpha}
    =
    \int
    x\,
    \boldsymbol{\Psi}^\dagger(x,t)
    \sigma_\alpha
    \boldsymbol{\Psi}(x,t)
    d x,
    \qquad
    \alpha=y,z.
\end{equation}
Since $x$ and $\sigma_\alpha$ act on different factors of the Hilbert space, they commute and no additional symmetrization is required. To separate width effects from the mass variation inherent in the fixed-$m\sigma^2$ sequence, an additional control calculation is performed at fixed $m=1$ and fixed initial velocity $v_0=-1$, while $\sigma=0.15,0.2,0.25$ and $0.50$ is varied. All other quantum propagation parameters are identical to those of the main Rabi calculation. The coupling-strength scan uses $g_R=0.01,0.1,1$ and $2$, with $\sigma=0.25,m=1$, and all other grid parameters unchanged. The semiclassical equations are integrated with the same initial mean position, momentum, and spin state as the full quantum calculation.

\subsection{Internal-state-dependent force model}

For the ISDF model, the spatial density and Bohmian velocity are evaluated analytically; no numerical Schr\"{o}dinger propagation is required. We use $E_g=0,E_e=1,f=0.25,x_0=p_0=0$, and $P_g=1/3, P_e=2/3$ (or other sets of $P_g,P_e$). For each $\sigma$, $N=3000$ initial positions are sampled from the Gaussian density and propagated to $t_{max}=20$ with a fourth-order Runge–Kutta step of 0.02. In regions where the analytical density falls below $10^{-14}$, the numerical velocity is set to zero; sampled trajectories do not enter these regions in the reported calculations. The trajectories are not clipped to a spatial grid. Dynamically enlarged spatial intervals are used only for density plots and analytical boundary-tail diagnostics.

The yellow and blue subensembles in Fig. \ref{fig:internal_preliminary} in the main text are classified according to the sign of $x(t_{max})$. They are therefore postselected spatial groups rather than preassigned internal-state branches.

%At each time, the numerical trajectory mean is compared with the exact Born expectation value \eqref{eq:internal_exact_mean}. The sampled position histogram is also compared with the analytical density in Eq. \eqref{eq:probdensity}. 
The branch-resolution parameter is evaluated directly from
\begin{equation}
    R(t)
    =
    \frac{ft^2}{2m\sigma(t)}.
\end{equation}
The spatial domain used for any auxiliary grid evaluation is chosen such that both accelerating branch centers and several time-dependent standard deviations remain inside the numerical box throughout the propagation.

\subsection{Convergence checks}
Three numerical checks are performed. First, the total quantum norm is monitored throughout all grid-based propagations. Second, the Bohmian ensemble average is compared with the Born expectation value; their difference decreases when the number of trajectories is increased. Third, the spatial box and grid resolution are enlarged until the probability density near the boundaries and the changes in the plotted observables become negligible. For the hard-core entropy calculation, convergence is additionally checked with respect to both the singular-value grid size and the entropy sampling interval.

% ============================================================
\section{Hard-core wavefunction and exact reduced probability flow}
\label{app:hard_core_derivation}
% ============================================================
In this section, we provide a detailed derivation of the system wavefunction in the hard-core model.

For two equal-mass particles of mass $m$, define the center-of-mass and relative coordinates
\begin{equation}
    X=\frac{x_1+x_2}{2},
    \qquad
    r=x_1-x_2.
    \label{eq:app_X_r}
\end{equation}
The kinetic energy separates as
\begin{equation}
    -\frac{\hbar^2}{2m}
    \left(
    \frac{\partial^2}{\partial x_1^2}
    +
    \frac{\partial^2}{\partial x_2^2}
    \right)
    =
    -\frac{\hbar^2}{4m}
    \frac{\partial^2}{\partial X^2}
    -
    \frac{\hbar^2}{m}
    \frac{\partial^2}{\partial r^2}.
    \label{eq:app_kinetic_separation}
\end{equation}
Equivalently, the center-of-mass coordinate has total mass $M=2m$, while the relative coordinate has reduced mass $\mu=m/2$. The hard-core condition $x_1=x_2$ becomes
\begin{equation}
    r=0,
\end{equation}
and the physically allowed region is $r>0$. The wavefunction therefore must satisfy
\begin{equation}
    \Psi(r=0,t)=0.
    \label{eq:app_boundary}
\end{equation}

When the initial widths are equal, the initial two-particle Gaussian factorizes into a center-of-mass Gaussian and a relative-coordinate Gaussian. The center-of-mass part evolves freely. The relative part evolves as a free packet on the half-line $r>0$ with a hard-wall boundary at $r=0$. This is obtained by the method of images:
\begin{equation}
    \phi_{HC}(r,t)
    =
    \phi_{\rm free}(r,t)
    -
    \phi_{\rm free}(-r,t),
    \qquad r>0.
    \label{eq:app_method_images}
\end{equation}
The odd image extension of the free relative-coordinate packet enforces the Dirichlet boundary condition
\begin{equation}
    \phi(0,t)=0.
\end{equation}
Combining the free center-of-mass evolution with Eq.~\eqref{eq:app_method_images} gives Eq.~\eqref{eq:hard_core_solution} in the main text.

To probe the exact reduced dynamics of particle 1 at the level of the reduced density and current, we integrate out $x_2$ to obtain the marginal density and current,
\begin{equation}
    \rho_1(x,t)
    =
    \int_{-\infty}^{x}
    |\Psi(x,x_2,t)|^2
    \,d  x_2,
    \label{eq:exact_marginal_density_hc}
\end{equation}
and
\begin{equation}
    J_1(x,t)
    =
    \int_{-\infty}^{x}
    \frac{\hbar}{m}
    \operatorname{Im}
    \left[
    \Psi^*(x,x_2,t)
    \frac{\partial\Psi(x,x_2,t)}{\partial x}
    \right]
    \,d  x_2.
    \label{eq:exact_marginal_current_hc}
\end{equation}
The contact term at $x_1=x_2$ vanishes because $\Psi(x_1=x_2,t)=0$, so no probability flows through the boundary. The corresponding reduced probability-flow velocity is $u_1(x,t)=J_1(x,t)/\rho_1(x,t)$, or
\begin{equation}
    u_1(x,t)=\frac{\hbar}{m}\frac{\int_{-\infty}^x v_1(x,x_2,t)|\Psi(x,x_2,t)|^2 d  x_2}{\int_{-\infty}^x |\Psi(x,x_2,t)|^2 d  x_2}
\end{equation}
It is a conditional average of the full configuration-space velocity at fixed $x_1=x$, not the velocity of an individual full two-body Bohmian trajectory. It reproduces the exact marginal density and mean position, but it is not generally identical to the actual Bohmian velocity of particle 1 along a full two-body trajectory. The continuity equation of the reduced density and current obeys
\begin{equation}
    \frac{\partial \rho_1}{\partial t}+\frac{\partial J_1}{\partial x}=0.
    \label{eq:reducedcont}
\end{equation}

On the other hand, this exact reduced dynamics is generally not equivalent to a closed one-particle Schr\"odinger equation with a local scalar potential. The normalized conditional potential may be denoted by 
\begin{equation}
    V_{cond}(x,t)=\frac{\int_{-\infty}^x V(x,x_2)|\Psi(x,x_2,t)|^2 d  x_2}{\int_{-\infty}^x |\Psi(x,x_2,t)|^2 d  x_2}.
    \label{eq:Vcond}
\end{equation}
This quantity is the conditional expectation value of the pair interaction at a fixed position $x$ of particle-1. It is not a genuine external potential; it depends on the full two-body state and generally does not yield Eq. \eqref{eq:reducedcont} when inserted in a one-body Schr\"{o}dinger equation. In the hard-core case, most of the conditional probability lies at positive separations for which the finite repulsive regularization is small, so the effective $V_{cond}$ is appreciable only when $x$ lies near the excluded region, but its exact form is nonlocal and time-dependent.

% ============================================================
\section{Additional figures}
\label{app:morefigures}
% ============================================================
In this section, we provide more figures supplementary to the plots in the main text.

\begin{figure}[H]
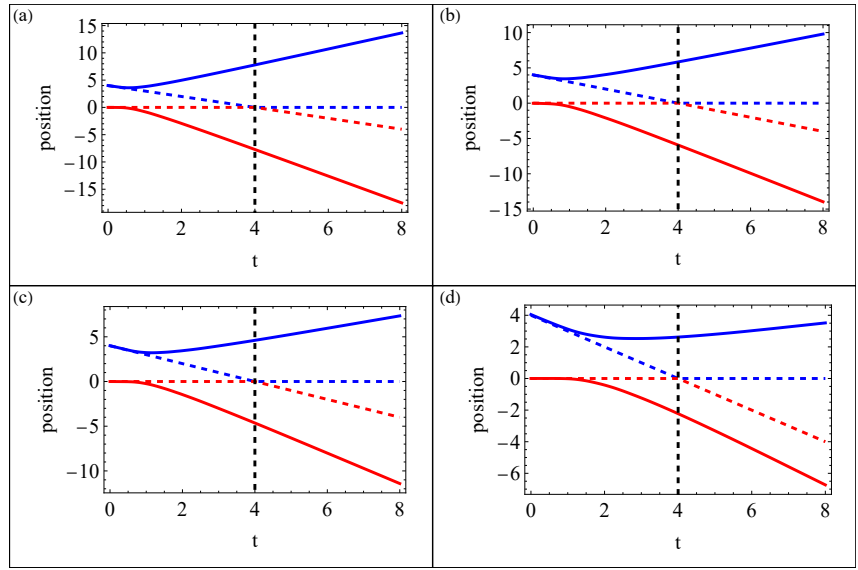

\centering
\safeincludegraphics[width=0.68\textwidth]{figs1.pdf}
\caption{Trajectories of two distinguishable equal-mass particles with a hard-core interaction. Panels (a)-(d) compare Bohmian ensemble averages, shown by solid lines, with the corresponding classical elastic-collision trajectories, shown by dashed lines. The initial Gaussian widths are $\sigma=0.15,0.20,0.25$, and $0.50$, while $m=1$ remains fixed. The initial conditions are $x_{1,0}=4$, $v_1=p_1=-1$, and $x_{2,0}=0$, $v_2=p_2=0$. Blue and red curves denote particles 1 and 2, respectively. The vertical black dashed line marks the common classical collision time $(t_c=4)$. Each Bohmian average is computed from 300 realizations.}
\label{fig:s1}
\end{figure}

\begin{figure}[H]
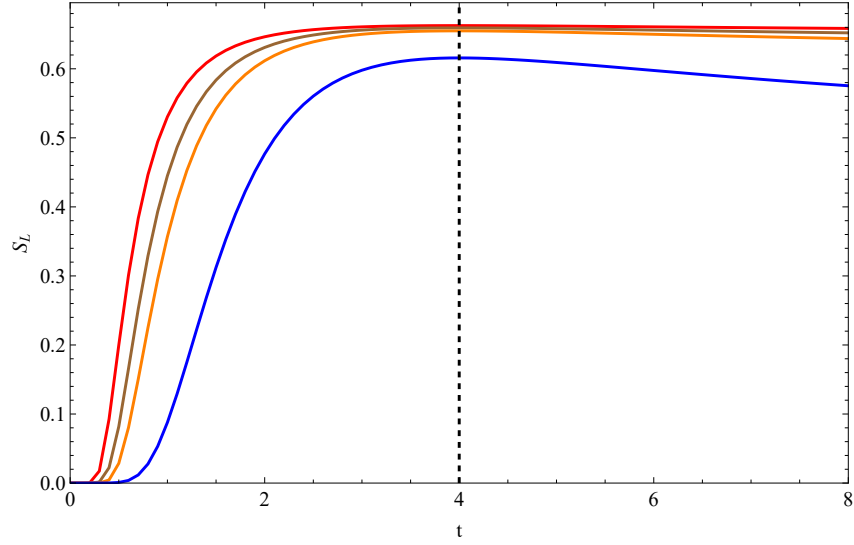

\centering
\safeincludegraphics[width=0.68\textwidth]{figs2.pdf}
\caption{The linear entropy, cf. Eq. \eqref{eq:linear_entropy_app} in the main text, for the hard-core collision, corresponding to the trajectory calculations in Fig.~\ref{fig:s1}. Red, brown, orange and blue curves correspond to $\sigma=0.15,0.20,0.25$ and $0.5$, respectively, with fixed $m=1$. The same fixed velocities and common collision time $t_c=4$ (marked by the vertical black dashed line) are used as in Fig. \ref{fig:s1}.}
\label{fig:s2}
\end{figure}

Figure~\ref{fig:s1} tracks the exact two-body hard-core motion for a fixed mass (and initial velocity), but different initial widths, and Fig. \ref{fig:s2} compares the corresponding linear entropy.

\begin{figure}[H]
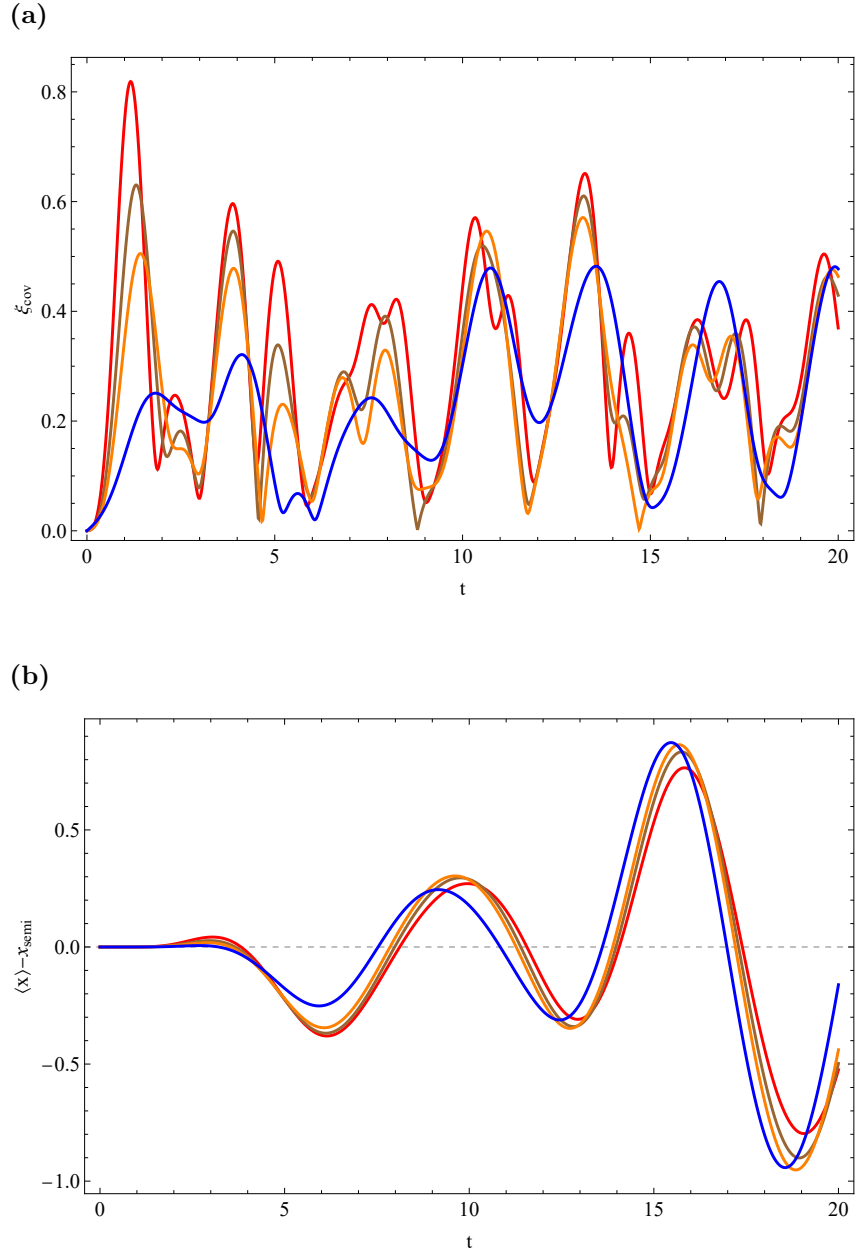

\centering
\subfloat[]{%
\safeincludegraphics[width=0.68\textwidth]{figs3a.pdf}
}
\\
\subfloat[]{%
\safeincludegraphics[width=0.68\textwidth]{figs3b.pdf}
}
\caption{Fixed-mass control of the Rabi-model closure error. The oscillator mass and initial velocity are fixed at $m=1$ and $v_0=-1$, respectively, while the initial Gaussian width is varied. Panel (a) shows the dimensionless covariance-induced closure-error measure $\xi_{cov}$ defined in Eq. \eqref{eq:spin_position_covariance} in the main text. Panel (b) shows the quantum–Ehrenfest position difference.  Red, brown, orange, and blue curves correspond to $\sigma=0.15,0.2,0.25$ and $0.5$, respectively. Narrower packets generate sharper early-time covariance peaks, whereas the broader packet exhibits more persistent late-time correlations and the largest late-time displacement error. Other parameters are $\hbar=\Omega=\omega=1$, $g_R=0.25,x_0=2$, and the initial spin state is $|\uparrow\rangle$.}
\label{fig:s3}
\end{figure}

\begin{figure}[H]
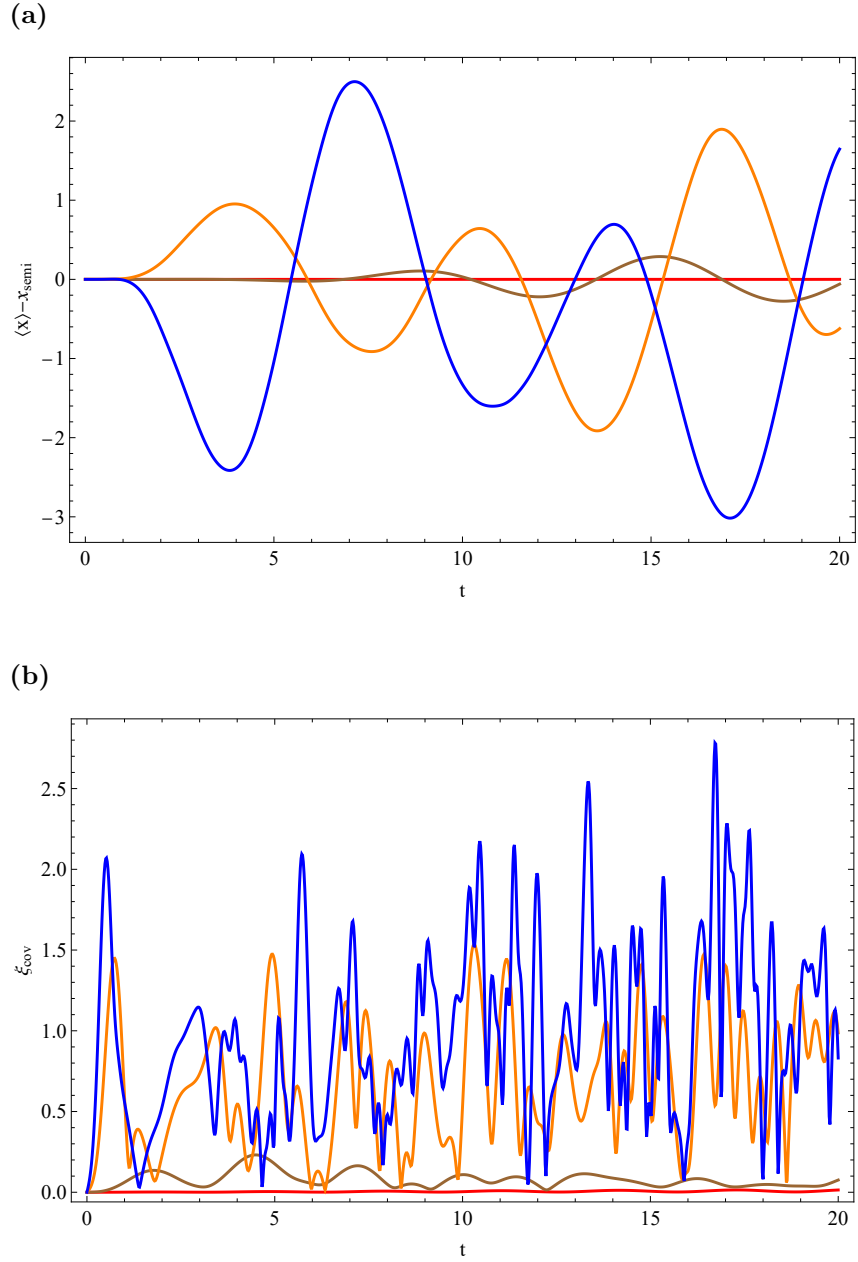

\centering
\subfloat[]{%
\safeincludegraphics[width=0.68\textwidth]{figs4a.pdf}
}
\\
\subfloat[]{%
\safeincludegraphics[width=0.68\textwidth]{figs4b.pdf}
}
\caption{Effect of increasing coupling on the Rabi-model closure error. Panel (a) measures the deviation $\langle x(t)\rangle-x_{semi}(t)$ between the quantum oscillator mean and its mixed semiclassical trajectory. Panel (b) shows the dimensionless covariance-induced closure-error measure $\xi_{cov}(t)$, cf. Eq. \eqref{eq:spin_position_covariance} in the main text. Red, brown, orange and blue lines correspond to coupling strength $g_R=0.01,0.1,1$ and $2$, with $\sigma=0.25,m=1$, and all other parameters as Fig. \ref{fig:rabi_trajectory}.}
\label{fig:s4}
\end{figure}

Figures~\ref{fig:s3} and \ref{fig:s4} measure the deviation of the Born expected position from the semiclassical trajectory and covariance-induced closure error for a varying $\sigma$, and for different spin-particle coupling strengths in the Rabi model, respectively, with fixed mass and (initial) velocity.

\begin{figure}[H]
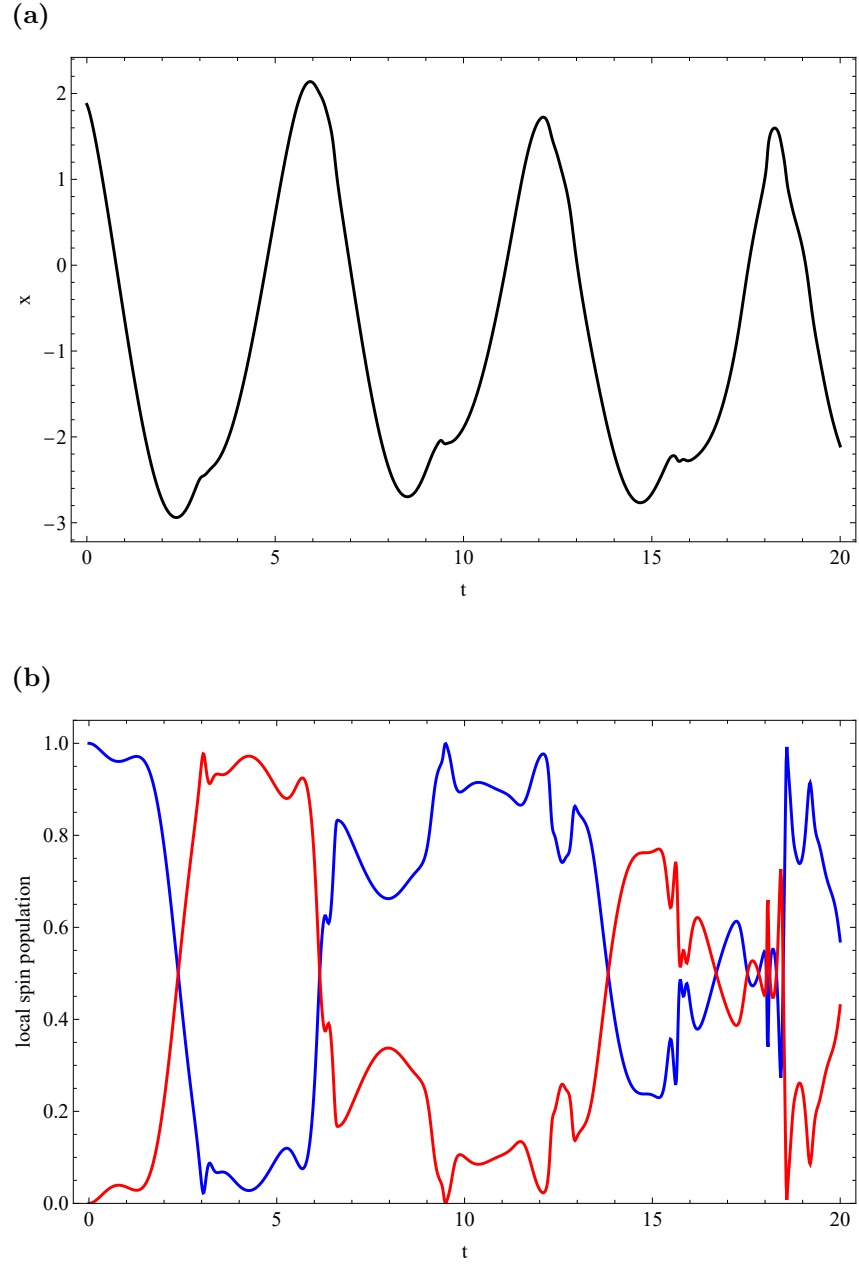

\centering
\subfloat[]{%
\safeincludegraphics[width=0.68\textwidth]{figs5a.pdf}
}
\\
\subfloat[]{%
\safeincludegraphics[width=0.68\textwidth]{figs5b.pdf}
}
\caption{Single-trajectory diagnostics in the Rabi model. Panel (a) shows one Bohmian trajectory selected from the ensemble in Fig.~\ref{fig:rabi_trajectory}(c). Panel (b) shows the corresponding local spinor weights evaluated along that trajectory according to Eq.~\eqref{eq:local_spin_population}.}
\label{fig:s5}
\end{figure}

Figure \ref{fig:s5} shows the evolution of local spinor weights evaluated along an individual Bohmian trajectory, 
\begin{equation}
    P_\alpha^{\rm loc}(t)
    =
    \frac{
    |\Psi_\alpha(x(t),t)|^2
    }{
    \sum_{\beta=1}^2|\Psi_\beta(x(t),t)|^2
    },
    \label{eq:local_spin_population}
\end{equation}
along a particular Bohmian trajectory.

%Figure \ref{fig:s5} shows the evolution of the internal state population in the ISDF model. 

\begin{figure}[H]
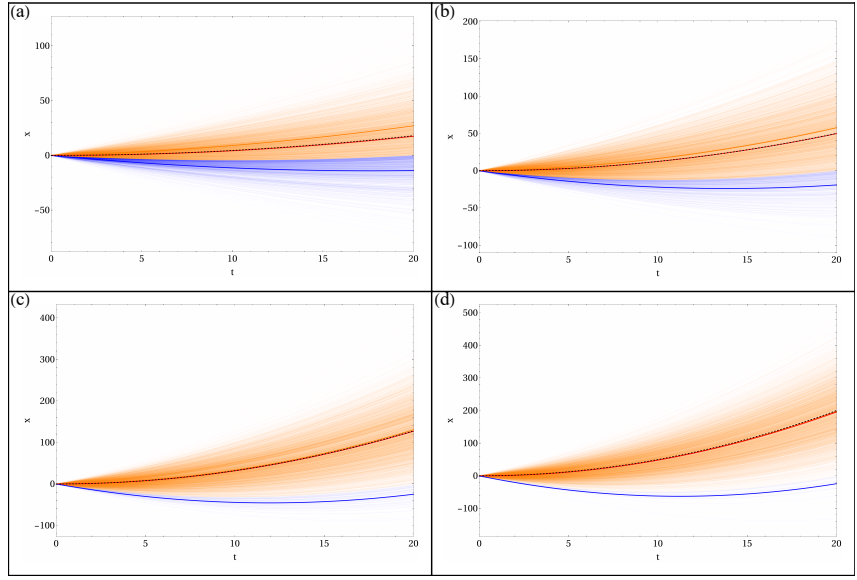

\centering
\safeincludegraphics[width=0.68\textwidth]{figs6.pdf}
\caption{Bohmian trajectories for the ISDF model with $P_g=0$ and $P_e=1$, namely purely excited in state $|e\rangle$ . Other parameters and format are the same as Fig. \ref{fig:internal_preliminary} in the main text.}
\label{fig:s6}
\end{figure}

\begin{figure}[H]
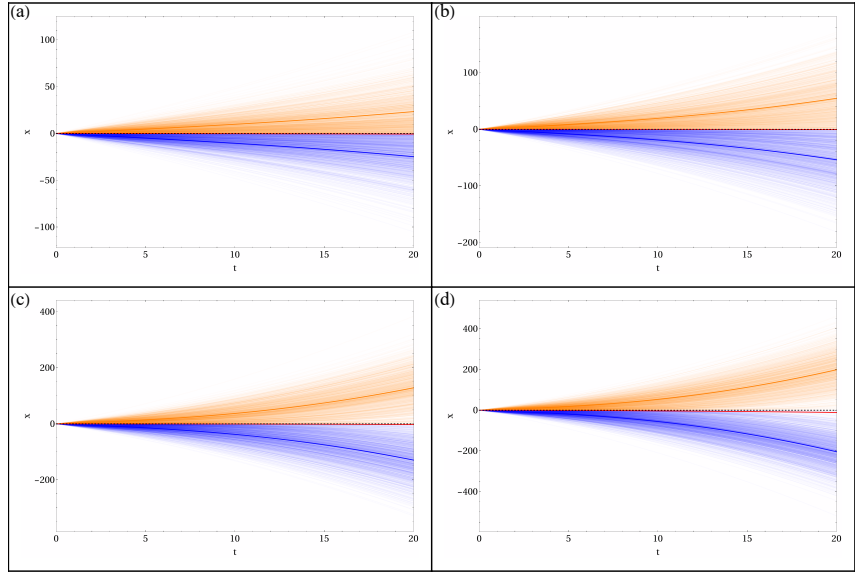

\centering
\safeincludegraphics[width=0.68\textwidth]{figs7.pdf}
\caption{Bohmian trajectories for the ISDF model with equal superposition, i.e., $P_g=P_e=1/2$. Other parameters and format are the same as Fig. \ref{fig:internal_preliminary} in the main text.}
\label{fig:s7}
\end{figure}

Figures \ref{fig:s6} and \ref{fig:s7} collect Bohmian trajectories of the ISDF model with respect to different initial widths $\sigma$ and internal states, with $P_g=0,P_e=1$ for Fig. \ref{fig:s6} and $P_g=P_e=1/2$ for Fig. \ref{fig:s7}, respectively.

%\begin{figure}[H]
%\centering
%\safeincludegraphics[width=0.68\textwidth]{figs5.pdf}
%\caption{Evolution of internal state populations in Fig. \ref{fig:internal_preliminary}. Blue and red lines refer to populations of internal states $|g\rangle$ and $|e\rangle$, respectively.}
%\label{fig:s5}
%\end{figure}

\section*{Acknowledgments} 
B.C. thanks Abraham Nitzan and Xiangrong Wang for insightful discussions. B.C. acknowledges the financial support of the National Natural Science Foundation of China (No. 12404232), start-up funding from the Chinese University of Hong Kong, Shenzhen (No. UDF01003468) and the Shenzhen city “Pengcheng Peacock” Talent Program.

% ============================================================
% Bibliography
% ============================================================

\bibliography{reference}
\end{document}